\documentclass[universe,article,accept,pdftex,moreauthors]{Definitions/mdpi}
\graphicspath{{figs/}}
\firstpage{1} 
\pubvolume{11}
\issuenum{5}
\articlenumber{158}
\pubyear{2025}
\copyrightyear{2025}
\externaleditor{Denis Leahy, Vlasios Petousis, Charalampos Moustakidis and Martin Veselsky} 
\datereceived{3 March 2025} 
\daterevised{7 May 2025} 
\dateaccepted{8 May 2025} 
\datepublished{12 May 2025} 
\hreflink{https://doi.org/10.3390/\linebreak universe11050158} 

\usepackage{algorithm}
\usepackage{algpseudocode}

\Title{Screening Mechanisms on White Dwarfs: Symmetron \linebreak and Dilaton} 

\TitleCitation{Screening Mechanisms on White Dwarfs: Symmetron and Dilaton}

\Author{{Joan} 
 Bachs-Esteban $^{1,}$*\orcidA{}, Ilídio Lopes $^{1}$\orcidB{} and Javier Rubio $^{2}$\orcidC{}}

\AuthorNames{Joan Bachs-Esteban, Ilídio Lopes and Javier Rubio}

\isAPAStyle{%
       \AuthorCitation{Lastname, F., Lastname, F., \& Lastname, F.}
         }{%
        \isChicagoStyle{%
        \AuthorCitation{Lastname, Firstname, Firstname Lastname, and Firstname Lastname.}
        }{
        \AuthorCitation{{Bachs-Esteban}
, J.; Lopes, I.; Rubio, J.}
        }
}

\address{%
$^{1}$ \quad Centro de Astrofísica e Gravitação---CENTRA, Departamento de Física, Instituto Superior Técnico---IST, Universidade de Lisboa---UL, Av. Rovisco Pais 1, 1049-001 Lisboa, Portugal; ilidio.lopes@tecnico.ulisboa.pt\\
$^{2}$ \quad {Departamento de Física Teórica and Instituto de Física de Partículas y del Cosmos} 
 (IPARCOS), Facultad de Ciencias Físicas, Universidad Complutense de Madrid, 28040 Madrid, Spain; javier.rubio@ucm.es}

\corres{Correspondence: joan.bachs@tecnico.ulisboa.pt}

\abstract{{This work provides the first comparison of the symmetron and dilaton fields in white dwarfs. We show how these screening} mechanisms behave inside {such stars} and their impact on stellar properties. Employing a custom-developed shooting method, we solve the scalar--tensor equilibrium equations in the Newtonian approximation. We consider a Chandrasekhar equation of state and examine a range of potential mass scales and coupling strengths for both fields.
Both fields enhance the pressure drop in low-density white dwarfs, leading to smaller stellar masses, radii, and luminosities. {Unlike chameleon models,} their effects are suppressed in more massive stars, with symmetron fields fully decoupling and dilaton fields weakening but not vanishing. Consequently, no mass--radius curve for screened white dwarfs exceeds the Newtonian prediction {in any of these three mechanisms.} The mass--radius deviations are generally more pronounced at lower densities, depending on model parameters.
Due to their common runaway potential, we confirm that dilaton and chameleon fields display similar field and gradient profiles. In contrast, due to their environment-dependent coupling, the dilaton and symmetron mechanisms exhibit stronger density-dependent screening effects.
These findings highlight both phenomenological differences and theoretical similarities among these mechanisms, motivating {asteroseismology studies to constrain the symmetron and dilaton \mbox{parameter spaces.}}}


\begin{document}




\section{\label{sec:intro}Introduction}

Among the various extensions of General Relativity (GR), scalar--tensor (ST) theories of gravity~\cite{Burrage_2018, Brax:2021wcv,Fischer:2024eic} stand out as simple and elegant frameworks. These theories generally introduce one or more scalar fields that can mediate an additional force---often referred to as a fifth force---between matter components. If sufficiently long-range, such a force could conflict with local gravitational tests~\cite{Will:2018bme} or impact the formation of structures in both the early~\cite{Amendola:2017xhl, Savastano:2019zpr,Goh:2023mau} and late universe~\cite{Wetterich:1994bg, Wetterich:2007kr,Amendola:2007yx, Casas:2016duf} (with notable exceptions discussed in~\cite{Garcia-Bellido:2011kqb, Ferreira:2016kxi,Casas:2018fum, Copeland:2021qby}). Many ST theories incorporate screening mechanisms to reconcile these potential issues, making the scalar field's effects dependent on the environment. For example, in the symmetron~\cite{Hinterbichler_2010} and dilaton mechanisms~\cite{Brax_2010}, the coupling to matter varies with environmental conditions, while in the chameleon scenario~\cite{Khoury_2004a}, the scalar field's mass changes.~{Vainshtein~\cite{Vainshtein_1972} screening---which can be exhibited by massive Galileons~\cite{Burrage_2021}---depends on the non-linear interaction terms becoming dominant in the vicinity of compact sources. Other models rely on the variability of the kinetic function, like k-mouflage~\cite{Babichev_2009, Brax_2016} and k-essence~\cite{Brax_2013, ter_Haar_2021}, which can be extended to generalised k-essence~\cite{Luongo:2024opv} and quasi-quintessence~\cite{D_Agostino_2022}}. These screening mechanisms ensure that the scalar field remains negligible on astrophysical scales but can still be relevant on cosmological scales. 

\textls[-15]{The partial breakdown of these screening mechanisms within massive objects is expected to influence the equilibrium structure of stars, thereby affecting fundamental properties such as mass--radius relations and cooling times (see, e.g.,~\cite{Babichev_2009, Chang:2010xh,Sakstein_2013, Brito:2014ifa,Babichev:2016jom, deAguiar_2020, deAguiar_2021, Panotopoulos:2021rih,ter_Haar_2021, Dima_2021}). Among the various compact objects, white dwarfs (WDs) are particularly promising yet underexplored targets for studying the effects of screening mechanisms~\cite{Saltas_2018, Alam_2023, Kalita_2023}. This is primarily due to two factors: first, the equation of state (EoS) governing the matter inside WDs is well understood~\cite{Camenzind_2007}, and second, the wealth of observational data available from \textit{{Gaia}
} data releases~\cite{Jim_nez_Esteban_2018, Tremblay_2018, Kilic_2020} provides extensive {and precise} information on the spatial distribution, kinematics, and key properties of {about $10^5$ WD candidates}, such as luminosity, temperature, and radius.}

In this study, we focus on examining the effects of symmetron and dilaton fields on the structure of WDs, {both of which are scalar fields whose screening mechanism depends on their environment-dependent coupling. The difference between them lies in how they couple to matter: the symmetron is built on symmetry restoration in dense regions, while the dilaton's coupling strength is inversely proportional to the environment density. In contrast, the chameleon's coupling strength is constant, with its effective mass regulating the field's interaction. This is why chameleon-screened WDs exhibit a universal deviation from Newtonian results across stars with different masses and radii~\cite{bachsesteban2024}. In the present work, we show that symmetrons decouple completely in the most dense (and more massive) stars, a behaviour also displayed by dilatons as a function of their parameters.} 
Since WDs can be adequately described as non-relativistic objects both in GR and the ST scenarios considered here, we numerically solve the equilibrium equations in the Newtonian approximation utilising a Chandrasekhar EoS. In addition, we investigate how the scalar field influences the pressure, specific heat, and luminosity of the stars and analyse the field profiles. We derive the corresponding mass--radius curves by exploring various potential scales and coupling strengths to matter. 

This paper is organised as follows: Section \ref{sec:theory} introduces the theoretical framework, discussing the ST theory together with the equilibrium equations for static and spherically symmetric WDs and their EoS. Section \ref{sec:Model} describes the specific screening mechanisms under consideration and outlines the numerical methods employed, including boundary conditions and our customised shooting method. Section \ref{sec:Results} presents our findings on symmetron- and dilaton-screened WDs, followed by a discussion of their implications and potential future research directions in Section \ref{sec:Conclusion}. In this work, we use the metric signature ($-$, +, +, +) and consider $c=\hbar=k_B=1$, unless otherwise stated.

\section{\label{sec:theory}Framework}

\subsection{\label{subsec:ST} Scalar--Tensor Theory}

Several scalar--tensor (ST) theories featuring environmentally dependent screening mechanisms, such as symmetrons, dilatons, or chameleons, can be described by the following general action~\cite{Sakstein_2013}:\vspace{-6pt}
\begin{equation}\label{eq:STaction}
    S=\int d^4x\sqrt{-g}\left[\frac{M_P^2}{2}R - \frac{1}{2}\nabla_\mu\phi\nabla^\mu\phi - V(\phi)\right] + S_m\left[\Psi_m;A^2(\phi)g_{\mu\nu}\right].
\end{equation}

{This} 
 action cannot accommodate screening mechanisms that work with the kinetic term, e.g., Vainshtein~\cite{Vainshtein_1972} or k-mouflage~\cite{Babichev_2009}, since those models have non-canonical kinetic functions. However, there are $f(R)$ theories and low-energy limits of string theories that can be recast into an action of this form~\cite{Bettoni_2022}. Here, $M_P=(8\pi G)^{-1/2}=2.43\times 10^{18}$ GeV is the reduced Planck mass, $g$ and $R$ denote the determinant and Ricci scalar of the \textit{{Einstein frame}} metric $g_{\mu\nu}$, and $\phi$ represents a scalar field. Each theory within this framework is characterised by a self-interacting potential $V(\phi)$ and a conformal coupling function $A(\phi)$, which governs the interaction between the scalar field and matter fields $\Psi_m$. Specifically, the scalar field is gravitationally coupled to the matter fields via a conformally rescaled \textit{{Jordan frame}} metric $\Tilde{g}_{\mu\nu}\equiv A^2(\phi)g_{\mu\nu}$. This coupling alters the Newtonian force in the non-relativistic limit, categorising these ST theories as modified gravity (MG) theories~\cite{Sakstein_2013}.
 
We obtain the field equations by varying the action \eqref{eq:STaction} with respect to the Einstein frame metric\vspace{-6pt}
\begin{equation}\label{eq:EF_eq}
    G_{\mu\nu}=\kappa^2\left[T_{\mu\nu}+\nabla_\mu\phi\nabla_\nu\phi-g_{\mu\nu}\left(\frac{1}{2}\nabla_\sigma\phi\nabla^\sigma\phi+V(\phi)\right)\right]\,,
\end{equation}
where $G_{\mu\nu}$ is the Einstein tensor, $\kappa\equiv M_P^{-1}$, and \vspace{-6pt}\begin{equation}\label{eq:EF_Tensor_Def}
    T_{\mu\nu}\equiv-\frac{2}{\sqrt{-g}}\frac{\delta S_m}{\delta g^{\mu\nu}}
\end{equation}
represents the energy--momentum tensor of the matter fields, which we assume to be perfect fluids, specifically
\begin{equation}\label{eq:EF_Tensor_Perfect_Fluid}
    T^{\mu\nu}\equiv(\epsilon+P)u^\mu u^\nu+Pg^{\mu\nu}\,,
\end{equation}
where $u^\mu$ denotes the four-velocity of the fluid elements, and $\epsilon$ and $P$ represent the total energy density and pressure in the fluid's rest frame, respectively. To obtain the equation of motion for the scalar field, we need to vary the action \eqref{eq:STaction} with respect to the {field}
\vspace{-6pt}\begin{equation}\label{eq:phi_eq}
    \Box\phi=\frac{dV(\phi)}{d\phi}-\frac{d\text{ ln }A(\phi)}{d\phi}T\equiv\frac{dV_{\rm eff}(\phi)}{d\phi}\,,
\end{equation}
where $V_{\rm eff}(\phi)$ is the effective potential that governs $\phi$. Finally, the matter equation of motion is determined by the divergence of Equation~\eqref{eq:EF_eq}:
\begin{equation}\label{eq:EF_Matter_EOM}
    \nabla^\nu T_{\mu\nu}=\frac{d\text{ ln }A(\phi)}{d\phi}T\nabla_\mu\phi\,,
\end{equation}
\textls[-5]{with $T\equiv g^{\mu\nu}T_{\mu\nu}$ denoting the trace of the energy--momentum tensor. This equation implies that particles do not follow geodesics in the Einstein frame metric $g_{\mu\nu}$; instead, their trajectories are also influenced by the gradient of the scalar field. Still, in an alternative but equally valid approach, we can then as well define the Jordan frame matter energy--momentum tensor as follows:}
\begin{equation}\label{eq:JF_Tensor_Def}
    \Tilde{T}_{\mu\nu}\equiv-\frac{2}{\sqrt{-\Tilde{g}}}\frac{\delta S_m}{\delta \Tilde{g}^{\mu\nu}}\,.
\end{equation}

{By} comparing this expression with Equation~\eqref{eq:EF_Tensor_Def}, it becomes clear that the two tensors are related by $T_{\mu\nu}=A^2(\phi)\Tilde{T}_{\mu\nu}$. Using Equation~\eqref{eq:EF_eq}, it can be shown that $\Tilde{T}_{\mu\nu}$ is indeed covariantly conserved, meaning $\Tilde{\nabla}^\nu\Tilde{T}_{\mu\nu}=0$, and that free particles follow geodesics determined by $\Tilde{g}_{\mu\nu}$. From the normalisation condition for the four-velocity, $g_{\mu\nu} u^\mu u^\nu =-1$, we derive the relation $u^\mu=A(\phi)\Tilde{u}^\mu$. This conformal transformation, along with the corresponding transformation for the energy--momentum tensor, allows us to establish the relationship between the fluid variables in both frames, specifically $\epsilon=A^4(\phi)\Tilde{\epsilon}$ and $P=A^4(\phi)\Tilde{P}$.

\subsection{\label{subsec:EoS}Equation of State}

The equation of state (EoS) encapsulates the microphysics of the stellar interior by relating pressure to density. As discussed here, WDs can be effectively described as non-relativistic objects in both GR and ST scenarios. Consequently, it is useful to introduce the rest-mass density $\Tilde{\rho}$ and the internal energy density $\Tilde{\Pi}$, which is related to the total energy density $\Tilde{\epsilon}$ through the expression\vspace{-6pt}
\begin{equation}\label{eq:rest_mass_density}
    \Tilde{\epsilon}\equiv \Tilde{\rho}\left(1+\frac{\Tilde{\Pi}}{\Tilde{\rho} c^2}\right)\,.
\end{equation}
where the speed of light $c$ is explicitly included to highlight that $\Tilde{\Pi}$ represents a first-order relativistic correction.

Furthermore, the pressure and energy in WDs are predominantly nonthermal, meaning that thermal effects can be treated as minor perturbations superimposed on the fluid dynamics~\cite{Camenzind_2007}. As a result, the EoS simplifies to single-parameter functions, specifically $\Tilde{P}(\Tilde{\rho})$ and $\Tilde{\epsilon}(\Tilde{\rho})$. It is important to note that we reference the EoS in terms of the Jordan-frame variables because the standard thermodynamic relation for energy conservation, $d(\Tilde{\epsilon}/\Tilde{\rho})=-\Tilde{P}d(1/\Tilde{\rho})$, is valid only in this frame.  

In WDs, the electrostatic energy of the matter structure is negligible compared to the Fermi energies, making Coulomb forces insignificant as well. Consequently, the electron pressure can be expressed as follows~\cite{Camenzind_2007}:
\begin{equation}\label{eq:Electron_pressure}
    \Tilde{P}=\frac{2}{(2\pi)^3 \hbar^3}\int_0^{p_{F,e}}\frac{p^2c^2}{\sqrt{p^2c^2+\left(m_ec^2\right)^2}}4\pi p^2dp=\frac{m_ec^2}{\lambda_e^3}\psi(x)\,,
\end{equation}
with\vspace{-6pt}
\begin{equation}\label{eq:Electron_pressure_formula}
    \psi(x)=\frac{1}{8\pi^2}\left[x\sqrt{1+x^2}\left(\frac{2x^2}{3}-1\right)+\text{ln}\left(x+\sqrt{1+x^2}\right)\right].
\end{equation}

{For} clarity, the factors $\hbar$ and $c$ are explicitly included here. The factor of 2 in Equation~\eqref{eq:Electron_pressure} accounts for the electron spin degeneracy, $p_F$ represents the Fermi momentum of the electrons, $m_e$ is the electron mass, $\lambda_e\equiv \hbar/(m_e c)$ is the electron Compton wavelength, and $x\equiv p_F/m_ec$ is the dimensionless Fermi momentum.

While the primary contribution to the pressure in WD comes from degenerate electrons, the energy is largely ion-dominated. Since the ions are non-relativistic at densities below the neutron drip threshold $\Tilde{\rho}_{\text{n-drip}}\simeq4\times10^{11}\text{ g}\,\text{cm}^{-3}$~\cite{Camenzind_2007}, the energy density can be approximated by the rest-mass density\vspace{-6pt}
\begin{equation}\label{eq:Newtonian_rho}
    \Tilde{\epsilon}=\Tilde{\rho}=\frac{m_Bn_e}{Y_e}\,,
\end{equation}
where $m_B$ is the mean nucleon mass, $n_e=8\pi p_{F,e}^3/(3h^3)$ is the electron number density, and $Y_e = Z/A$ (where $Z$ is the atomic number and $A$ is the atomic weight) is the mean number of electrons per nucleon. WDs are typically modelled as cold, degenerate matter composed of helium, carbon, or oxygen\endnote{The first EoS for such stars was derived by Chandrasekhar~\cite{Chandrasekhar:1935zz}. Hamada and Salpeter added zero-temperature corrections~\cite{1961ApJ...134..683H}.}. For these elements, $Y_e =0.5$ when fully ionised~\cite{Camenzind_2007}. The mean nucleon mass of carbon is $m_{B,C}=1.66057\times10^{-24}\,\text{g}$. Thus, the density $\Tilde{\rho}$ and pressure $\Tilde{P}$ can be expressed as follows:\vspace{-6pt}
\begin{equation}\label{eq:EoS}
    \Tilde{\rho}(x)=1.9479\times10^6\,x^3\,\text{g}\,\text{cm}^{-3},\quad\text{and}\quad\Tilde{P}(x)=1.4218\times10^{25}\,\psi(x)\,\text{dyn}\,\text{cm}^{-2}.
\end{equation}

Given the exploratory nature of this study and its efficiency in calculations, we chose the above equation for its simplicity. A more refined approach could involve the crystallisation process of WDs~\cite{Vidal:2024wto} as well as advanced EoS that consider magnetic effects~\cite{Das:2014ssa,Roy:2019nja}. Still, employing a more precise EoS would quantitatively change but not substantially affect our key results and conclusions. {For instance, the Hamada--Salpeter corrections~\cite{1961ApJ...134..683H} consider the Coulomb interactions between the electron gas and the cations as the primary correction to the free electron gas pressure. Thus, only pressure is affected by these considerations, while the density continues to be described in the non-relativistic way of Equation~\eqref{eq:Newtonian_rho}. Moreover, this modification to the pressure only implies a 5\% reduction in the radius for massive white dwarfs with the same mass~\cite{Saltas_2018}. Thus, the results would be very close to the ones presented here if we used that EoS. Adding dependence on temperature profiles or envelopes~\cite{Fontaine_2001} can decrease the radius of a certain star by 10\%, even up to 40\% for the less-dense stars~\cite{Saltas_2018}. However, as we will see in Section 4.4
, some realisations of the symmetron can lead to larger differences. Altogether, the Chandrasekhar EoS is sufficient for our purposes since we focus on the scalar fields.}

\subsection{\label{subsec:EqEq}Equilibrium Equations}

In the Newtonian framework, where the gravitational field is weak and static, and particles move slowly compared to the speed of light~\cite{Misner_1973}, pressure is negligible relative to density. For instance, for a massive WD with central density $\rho_0$$\sim$$10^{10} \text{ g}\,\text{cm}^{-3}$ {(the maximum value we consider in this work (see Section~\ref{sec:Results}) which yields the highest pressure)}, its core pressure is $P_0$$\sim$$10^{28} \text{dyn}\,\text{cm}^{-2}$, which translates to $P_0 / \rho_0$$\sim$$10^{-3}$ when we consider Planck units. Plus, we have tested the accuracy of the Newtonian description against the relativistic one, as we did in~\cite{bachsesteban2024}. As with chameleon fields, we have found that the masses and radii predicted by the Tolman–Oppenheimer–Volkoff equation agree with those calculated in the Newtonian limit for both the symmetron and dilaton scenarios. Consequently, the line element for a WD can be expressed as\vspace{-6pt}
\begin{equation}\label{eq:metric_spherical_polar_NewtonLimit}
    ds^2=-\left(1+2\Phi(r)\right)dt^2+\left(1-2\Phi(r)\right)dr^2+r^2d\Omega^2\,,
\end{equation}
where $\Phi(r)$ is the Newtonian potential and $d\Omega^2=d\theta^2+\text{sin}^2\theta d\varphi^2$ is the two-sphere line element. Substituting this metric into Equations~\eqref{eq:EF_eq}, \eqref{eq:phi_eq} and \eqref{eq:EF_Matter_EOM}, we obtain\vspace{-6pt}
\begin{eqnarray}
    \Phi'=&&\frac{m}{r^2}\,,\label{eq:background_Phi_Newton}\\
    m'=&&\frac{\kappa^2}{2}r^2A^4\Tilde{\rho}\,,\label{eq:background_m_Newton}\\
    \Tilde{P}'=&&-\Tilde{\rho}\left(\Phi'+\frac{A_{,\phi}}{A}\sigma \right)\,,\label{eq:background_p_Newton}\\
    \phi'=&&\sigma\,,\label{eq:background_phi_Newton}\\
    \sigma'=&&-\frac{2}{r}\sigma+V_{,\phi}+A_{,\phi}A^3\Tilde{\rho}\,,\label{eq:background_sigma_Newton}
\end{eqnarray}
where pressure contributions have been disregarded compared to energy contributions and second-order terms have been neglected. Additionally, $\Tilde{\epsilon}$ has been replaced with $\Tilde{\rho}$, as these two quantities are equivalent in the Newtonian limit (see Equation~\eqref{eq:rest_mass_density}).

After selecting the model functions $V(\phi)$ and $A(\phi)$, along with an appropriate EoS, the resulting ODE system can be numerically integrated starting from the origin. For each chosen central density $\Tilde{\rho}_0=\Tilde{\rho}(0)$, the integration yields the corresponding stellar mass $M$ and radius $R$. By repeating this process across a range of central densities, a mass--radius (MR) curve can be generated. This family of stars is uniquely determined by the EoS and is parameterised by the central density $\Tilde{\rho}_0$~\cite{Camenzind_2007}. Further details of the integration procedure and boundary conditions are provided in Section~\ref{subsec:BC}. {Note that this ODE system, as well as the entire formalism of the paper, is set in the Einstein frame. We consider density and pressure in the Jordan frame because the {EoS}
~\eqref{eq:EoS} is derived with no scalar field  present. Since $\rho$ and $P$ intrinsically depend on the scalar field, we cannot relate them through those expressions, we must use $\Tilde{\rho}$ and $\Tilde{P}$ instead. In addition, we point out that the mass $m$ that we integrate in Equation~\eqref{eq:background_m_Newton} is none other than the mass that we would observe from Earth, as indicated by the expression for the gravitational field that it generates, Equation~\eqref{eq:background_Phi_Newton}.}

\section{\label{sec:Model}Models}

\subsection{\label{subsec:Symm}Symmetron Screening}

Let us first discuss symmetron fields~\cite{Hinterbichler_2010}, scalar fields that feature a screening mechanism that relies on local symmetry restoration in the presence of matter. The coupling of a symmetron field to matter is proportional to its vacuum expectation value (VEV), which depends on the environment density. In low-density regions, the VEV becomes large and the symmetron couples with gravitational strength. Conversely, the symmetron's VEV falls to zero in high-density regions, leaving the scalar field screened and decoupled from matter. This is the result of the interaction between a symmetry-breaking potential and a universal coupling to matter, such as the following:\vspace{-6pt}
\begin{equation}\label{eq:sym}
    V(\phi)=-\frac{1}{2}\mu^2\phi^2+\frac{1}{4}\lambda\phi^4,\quad\text{and}\quad A(\phi)=1+\frac{1}{2}\frac{\phi^2}{M_S^2}+\mathcal{O}\left(\frac{\phi^4}{M_S^4}\right),
\end{equation}
which are defined by the mass scales $\mu$ and $M_S$ and the dimensionless coupling $\lambda$. The tachyonic mass term of the potential is responsible for the spontaneous breaking of the $\mathbb{Z}_2$ symmetry $\phi\rightarrow-\phi$. This choice of $V(\phi)$ and $A(\phi)$ means that the relevant field range is below the coupling mass scale $M_S$~\cite{Hinterbichler_2011}, that is $\phi\ll M_S$; hence, one can neglect any higher-order terms in the coupling function. In this way, the symmetron model naturally becomes an effective field theory with cutoff $M_S$. 

We can obtain the symmetron effective potential from Equation~\eqref{eq:phi_eq}, up to an integration constant irrelevant for its dynamics (see Equations~\eqref{eq:background_phi_Newton} and \eqref{eq:background_sigma_Newton}). In the Newtonian limit, pressure is negligible in front of density; hence, the trace of the energy--momentum tensor can be approximated by $T=3p-\rho\approx-\rho$ (see Section~\ref{subsec:EoS}). Since $\phi\ll M_S$, we can expand the coupling function logarithm $\text{ln }A(\phi)$ and write the effective potential as \vspace{-6pt}
\begin{equation}\label{eq:sym_eff_potential}
    V_{\rm eff}(\phi) \approx \frac{1}{2}\left(\frac{\rho}{M_S^2}-\mu^2\right)\phi^2+\frac{1}{4}\lambda\phi^4.
\end{equation} 

{One} can see from the expression above that, if the density is high enough such that $\rho>\mu^2M_S^2$, the symmetron VEV goes to zero---i.e., $\phi_{\text{VEV}}=0$---and the $\mathbb{Z}_2$ symmetry remains intact. If $\rho<\mu^2M_S^2$, the reflection symmetry is spontaneously broken and the symmetron field acquires a VEV whose value is $\phi_{\text{VEV}}=\mu/\sqrt{\lambda}$ in regions of very low density, such as large cosmic voids, where $\rho\simeq0$. {Consequently, the critical density $\rho_S\equiv\mu^2M_S^2$ can be regarded as the order parameter~\cite{Frohlich:1981yi} of the symmetron model.} Therefore, the fifth force mediated by the symmetron field will be comparable to gravity in a vacuum if $\phi_{\text{VEV}}/M_S^2\approx 1/M_P$~\cite{Hinterbichler_2011}, which translates to\vspace{-6pt}
\begin{equation}\label{eq:lambda_constraint}
    \frac{\mu}{M_S^2\sqrt{\lambda}}\approx\frac{1}{M_P}.
\end{equation} 

Let us consider a static, spherically symmetric WD of constant density $\rho_0$ surrounded by a medium whose density $\rho_\infty$ is much smaller than that of the star---i.e.,  $\rho_\infty\ll\rho_0$---for instance, the cosmological background. Assuming that $\rho_0>\mu^2M_S^2$ and that $\rho_\infty<\mu^2M_S^2$, the symmetron field will set to the value $\Bar{\phi}_0\approx0$ within the star and to $\Bar{\phi}_\infty\approx\mu/\sqrt{\lambda}$ outside of it. Both scalar field values are the minima of the effective potential in Equation~\eqref{eq:sym_eff_potential} in their respective regions. The mass of small fluctuations around these minima is  given by\vspace{-6pt}
\begin{equation}\label{eq:sym_eff_masses}
    m_0=\sqrt{\frac{\rho_0}{M_S^2}-\mu^2},\quad\text{and}\quad m_\infty=\sqrt{2}\mu,
\end{equation}
where the $\rho_0>\mu^2M_S^2$ condition is again necessary for the mass $m_0$ to be well  defined. 

{The transition between the scalar field value inside the WD and outside of it can be more or less abrupt (see {Figure}~4
). The symmetron can remain at $\Bar{\phi}_0$ for most of the star's interior and not start to vary until it is close to its radius $R$. This means that the change occurs in a shell of thickness $\Delta R$ at the outermost part of the star. Therefore,} analogous to the chameleon mechanism, one can define a \textit{{thin-shell}} factor for the symmetron one to discern between screened and unscreened objects~\cite{Hinterbichler_2011}, namely\vspace{-6pt}
\begin{equation}\label{eq:thin_shell_factor}
    \frac{\Delta R}{R}\equiv\frac{M_S^2}{6M_P^2\Phi}.
\end{equation}

{Astrophysical} bodies with a large (small) surface gravitational potential $\Phi$ will show a small (large) thin-shell factor---i.e., $\Delta R / R \ll 1$ ($\Delta R / R \gg 1$)---implying that the scalar field will be screened (unscreened) within them.

A typical WD with mass $M$$\sim$$0.5\,M_\odot$, radius $R$$\sim$$10^4$ km, and central density $\rho_0$$\sim$$10^6 \text{ g}\,\text{cm}^{-3}$ has a gravitational potential of $\Phi$$\sim$$10^{-4}$. The thin-shell factor \mbox{Relation \eqref{eq:thin_shell_factor}} tells us this star will be screened if $M_S \lesssim 10^{-2}\,M_P$. Imposing that the symmetron's interaction range is of the stellar radius, we can estimate the appropriate mass scale of the potential for a screened WD from Equation~\eqref{eq:sym_eff_masses}, which gives us $\mu$$\sim$$10^{-41}\,M_P$ for $M_S = 10^{-2}\,M_P$. For any mass scale pair $(M_S, \mu)$, the dimensionless coupling $\lambda$ is given by Equation~\eqref{eq:lambda_constraint}, \linebreak e.g., $\lambda$$\sim$$10^{-74}$ for the mass scales we have just calculated. We explore intervals around these estimates in the numerical results of Section~\ref{sec:Results}. We must bear in mind that applying these theoretical relations to a realistic WD gives us values that, despite being numerically valid, are incompatible with observational constraints~\cite{Hinterbichler_2011}. Yet, our goal here is to investigate the behaviour of WDs under symmetron gravity; this is why we must compromise with computational limitations, as customarily carried out for WDs and neutron stars~\cite{bachsesteban2024, Babichev:2010, Brax_2017, deAguiar_2020, Dima_2021, deAguiar_2021}.

\subsection{\label{subsec:Dila}Dilaton Screening}

The screening mechanism of the dilaton~\cite{Brax_2010} relies on its coupling function $A(\phi)$ minimising at $\phi_d$. Assuming that $\phi/M_P\ll1$, this means that $A(\phi)$ can be written as\vspace{-6pt}
\begin{equation}\label{eq:dil_coupling}
    A(\phi)=1+\frac{a_2}{2}\frac{\left(\phi-\phi_d\right)^2}{M_P^2}+\mathcal{O}\left(\frac{\phi^4}{M_P^4}\right),
\end{equation}
where $a_2$ is a positively defined dimensionless coupling. Thus, the dilaton coupling strength $\beta(\phi)\equiv M_P(\text{ln }A(\phi))_{,\phi}=a_2\left(\phi-\phi_d\right)/M_P$ becomes small when the field is near the minimum, effectively suppressing the fifth force. For the dilaton self-interaction potential $V(\phi)$, we consider the following runaway exponential:\vspace{-6pt}
\begin{equation}\label{eq:dil_potential}
     V(\phi)=A^4(\phi)V_0e^{-(\phi-\phi_d)/M_P},
\end{equation}
where $V_0$ is positive and has $M_P^4$ units. The effective potential for dilatons $V_{\rm eff}(\phi)$ derives from Equation~\eqref{eq:phi_eq}, as for the symmetrons. In the $\phi/M_P\ll1$ limit, we can expand the coupling function $A(\phi)$ and write $V_{\rm eff}(\phi)$ as 
\begin{equation}\label{eq:dil_eff_potential}
    V_{\rm eff}(\phi) \approx 
    V(\phi)-T\left[A(\phi)-1\right]\approx V(\phi)+\rho\left[A(\phi)-1\right],
\end{equation} 
where we have approximated the trace of the energy--momentum tensor by the rest-mass density (see Section~\ref{subsec:EoS}). In that same limit, both $A(\phi)$ and the exponential in $V(\phi)$ can be approximated by $1$, leading to the following estimate for the effective potential minimum:\vspace{-6pt}
\begin{equation}\label{eq:dil_effective_minimum}
    \Bar{\phi}-\phi_d\approx\frac{M_PV_0}{a_2\left(4V_0-T\right)}\approx\frac{M_PV_0}{a_2\left(4V_0+\rho\right)}.
\end{equation}

{The} above expression tells us that dilaton fields will be set to the value $\Bar{\phi}_\infty\approx M_P/(4a_2)+\phi_d$ in low-density regions---like the cosmological background or the interstellar medium, where $\rho\ll V_0$---which depends on the coupling function parameters. It also indicates that the dimensionless coupling must fulfil $a_2\gg1$ to be consistent with the $\phi/M_P\ll1$ assumption, which is what solar system tests entail~\cite{Brax_2010}. Inside a sufficiently dense body---like a static, spherically symmetric WD with constant density such that $\rho_0\gtrsim V_0$---the dilaton field will settle to the minimum $\Bar{\phi}_0$ given by Equation~\eqref{eq:dil_effective_minimum}. Since all parameters in that expression are positive, we realise that the dilaton field minimum inside the star is lower than outside, as happens for the symmetron and the chameleon mechanisms~\cite{bachsesteban2024}.

In the dilaton scenario, the mass of small fluctuations around a potential minimum $\Bar{\phi}$ corresponds to the value of the effective potential second derivative evaluated at $\Bar{\phi}$. Since $a_2\gg1$ and $\phi/M_P\ll1$, it can be estimated as
\begin{equation}\label{eq:dil_eff_mass_infty}
    m_{\rm eff}^2|_
    {\Bar{\phi}}\approx a_2\frac{4V_0+\rho}{M_P^2}
    =\frac{V_0}{M_P(\Bar{\phi}-\phi_d)}.
\end{equation}

{Then}, the dilaton effective mass within the WD $m_{\rm eff}^2|_{\Bar{\phi}_0}$---given by the above expression with $\rho_0$ instead of $\rho$---is bigger than the one outside of it $m_{\rm eff}^2|_{\Bar{\phi}_\infty}\approx 4a_2V_0/M_P^2$. This implies that the dilaton fifth force is short-range in dense environments and long-range on cosmological scales, illustrating the screening mechanism also exhibited by \mbox{chameleon fields.}

By imposing the interaction range of the dilaton field to be of the stellar radius, we can estimate the adequate potential scale for a screened WD. A radius of $R$$\sim$$10^4$ km and a central density $\rho_0$$\sim$$10^6 \text{ g}\,\text{cm}^{-3}$ are typical values for a WD. As we discussed earlier, the dimensionless coupling must be large. Hence, with $a_2 = 10^{2}$, one has that $V_0$$\sim$$10^{-85}\,M_P^4$. We explore intervals around these reference values in the numerical results of Section~\ref{sec:Results}. As before, these numerically valid values conflict with observational constraints. Still, we must accept computational limitations to investigate WDs under dilaton gravity, as carried out in previous studies (see Section~\ref{subsec:Symm}).

\textls[-15]{Comparing the coupling functions in Equations~\eqref{eq:sym} and \eqref{eq:dil_coupling}, we observe that they are almost identical for both scalar fields. This similarity is expected, as both screening mechanisms depend on the scalar field's coupling to matter, as discussed in Section~\ref{sec:intro}. The two couplings become effectively equivalent when setting $\phi_d=0$ in the dilaton model. The primary distinction lies in how the coupling strength is controlled: via the cutoff scale $M_S$ for the symmetron model and the dimensionless parameter $a_2$ for the dilaton scenario. Despite this resemblance, the self-interaction potentials differ significantly. The dilaton mechanism requires a runaway potential, like the chameleon one, but the exponential form in Equation~\eqref{eq:dil_potential} allows for negative scalar field values while maintaining a positive potential—something not possible in the chameleon model analysed in~\cite{bachsesteban2024}. Formally, the dilaton mechanism occupies an intermediate position between chameleon and symmetron scenarios, sharing the potential characteristics of the former and the coupling behaviour of the latter.}

\subsection{\label{subsec:BC}Boundary Conditions}

It is important to note that boundary conditions for the gravitational potential $\Phi(r)$ are not required, as long as the system is in equilibrium. This is because the system of Equations \eqref{eq:background_Phi_Newton}--\eqref{eq:background_sigma_Newton} depends solely on the radial derivatives of $\Phi(r)$. For the pressure, we set $\Tilde{p}(0)=\Tilde{p}_0$, where $\Tilde{p}_0$ represents the pressure at the centre of the WD. In our numerical integration, $\Tilde{p}_0$ will span a range of pressures within the validity domain of the EoS. Specifically, we will consider a range of central densities and compute the corresponding central pressures using the EoS described in Section~\ref{subsec:EoS} (see Equation~\eqref{eq:EoS}). The central densities we will use range from $\Tilde{\rho}_0 = 7\times10^{4}\text{ g}\,\text{cm}^{-3}$ to $\Tilde{\rho}_0 = 10^{10} \text{ g}\,\text{cm}^{-3}$, resulting in WDs with masses between $0.12$ and $1.42$ $M_\odot$ and radii from $1.3$ to $17.3$ km in Newtonian gravity, which are typical values for such stars. We set the initial condition $m(0)=0$ for the mass. Still, since our code starts from a small initial radius $r=r_0>0$, where the density is $\Tilde{\rho}_0$, the initial condition for the mass is $m(r_0)=(4/3)\pi r_0^3\Tilde{\rho}_0$.  

We assume that WDs are located in a galaxy with a density of $\Tilde{\rho}_G = 10^{-24} \text{ g}\,\text{cm}^{-3}$, meaning $\Tilde{\rho}$ equals $\Tilde{\rho}_G$ outside the star. This assumption is necessary to ensure that the spacetime transitions to Schwarzschild--de Sitter at large distances from the star. Consequently, the stellar radius $R$ is determined by the condition $\Tilde{\rho}(R)=\Tilde{\rho}_\infty=\Tilde{\rho}_G$. Yet, numerically, this translates to $\Tilde{\rho}(R)$ being approximately zero within a specified tolerance, as $\Tilde{\rho}_G$ is extremely small. In practice, we integrate up to a sufficiently large distance that is significantly greater than the typical WD radii and can be treated as effectively infinite. We identify the radial coordinate where the pressure falls below the tolerance, defining this as the stellar radius R. The stellar mass $M$ is then calculated as the mass enclosed within the sphere of radius $R$, as specified by Equation~\eqref{eq:background_m_Newton}.

The solution for the scalar field must remain regular at the centre of the star, which implies that the scalar field gradient satisfies $\sigma(0)=0$. While the value of the scalar field at the centre is unknown before integration, we do know that it will approach the exterior minimum at infinity, i.e., $\phi\rightarrow\Bar{\phi}_\infty$ as $r\rightarrow\infty$. This ensures that the solution remains regular at infinity, with $\sigma\rightarrow 0$ as $r\rightarrow\infty$. To solve this ODE system, we employ a shooting method to determine the correct value of  $\phi$ at $r=0$, yielding $\Bar{\phi}_\infty$ far from the star.

\subsection{\label{subsec:SM}Shooting Method}

\textls[-25]{The scalar field is governed by a second-order differential equation, Equation~\eqref{eq:phi_eq}, which we have reformulated into two first-order differential equations, \mbox{Equations~\eqref{eq:background_phi_Newton} and \eqref{eq:background_sigma_Newton}}.} As discussed in Section~\ref{subsec:BC}, the boundary conditions for the scalar field gradient $\sigma$ are known at both the centre and infinity, but for $\phi$, only the one at infinity is known, leaving the value at the origin to be determined.

We estimate the value of dilatons at the centre of the WD---let us call it $\phi_0$---from Equation~\eqref{eq:dil_effective_minimum}, while for symmetrons, we effectively have $\phi_0=0$. This is possible because WDs are Newtonian astrophysical objects and, as we explained in Sections~\ref{subsec:Symm} and~\ref{subsec:Dila}, both effective potentials have a minimum inside the star. We then numerically integrate the system of ODEs given by Equations~\eqref{eq:background_Phi_Newton}--\eqref{eq:background_sigma_Newton}. Following this, we calculate the relative error between the scalar field minimum at infinity, $\Bar{\phi}_\infty$, and the value obtained from our code, $\phi(r_{\text{max}})$, where $r_{\text{max}}$ is the maximum radial coordinate. If the relative error is smaller than the specified tolerance---i.e., if $|\Bar{\phi}_\infty - \phi(r_{\text{max}})| / \Bar{\phi}_\infty < \phi_{\text{tol}}$---then convergence has been achieved, and we store the result.

If the tolerance criterion is not satisfied, we adjust $\phi_0$ by a small increment $\delta\phi$, either increasing or decreasing it based on whether the difference between the theoretical and computed values, i.e., $\Bar{\phi}_\infty - \phi(r_{\text{max}})$, is positive or negative. At each iteration of the shooting method, we check the sign of this difference. When the sign changes (indicating that we have crossed the target value), we reduce $\delta\phi$. This approach accelerates convergence and improves precision with fewer steps. {Moreover, if the code stagnates or the convergence to the required tolerance $\phi_{\text{tol}}$ becomes too slow, it increases $r_{\text{max}}$ so that the scalar field has more space to evolve. This also means that we do not impose a strong integration limit, forcing the field to converge within a short distance. We have included a pseudocode for our shooting method in Appendix~\ref{sec:Pseudocode}.}

\section{Results}\label{sec:Results}

In this section, we present the results from numerically integrating \mbox{Equations~\eqref{eq:background_Phi_Newton}--\eqref{eq:background_sigma_Newton}}, using the model functions $V(\phi)$ and $A(\phi)$ from Equations~\eqref{eq:sym}, \eqref{eq:dil_coupling} and \eqref{eq:dil_potential} for the symmetron and dilaton mechanisms, respectively, and incorporating the EoS described by \mbox{Equation~\eqref{eq:EoS}}.~The integration is performed via the shooting method outlined in \linebreak Section~\ref{subsec:SM}. We considered central densities ranging from $\Tilde{\rho}_{0,\text{min}} = 7\times10^{4} \text{ g}\,\text{cm}^{-3}$ to $\Tilde{\rho}_{0,\text{max}} = 10^{10} \text{ g}\,\text{cm}^{-3}$, with a background density of $\Tilde{\rho}_\infty=2\times10^{-9}\text{ g}\,\text{cm}^{-3}$. For each central density, we computed the WD's radius $R$ and mass $M$ (see Section~\ref{subsec:EqEq}).

\textls[+25]{We have used the symmetron cutoff} $M_S=10^{-2}\,M_P$, mass scales in the \linebreak $\mu = 1 - 5 \times 10^{-41}\,M_P$ range, and the corresponding dimensionless couplings $\lambda$ according to \mbox{Relation \eqref{eq:lambda_constraint}}. For the dilaton mechanism, we set the minimum of the coupling function $A(\phi)$ to $\phi_d = 0$, equivalent to rescaling the field as $\phi-\phi_d\rightarrow\phi$. We consider dimensionless couplings between $a_2=10^{1}$ and $a_2=10^{4}$ and potential factors in the range $V_0=10^{-85} - 10^{-88}\,M_P^4$. 

Let it be noted that local gravity tests require $M_S\lesssim10^{-4}\,M_P$~\cite{Hinterbichler_2011}, and the potential mass scale should be $\mu\gtrsim10^{-56}\,M_P$ if we want symmetrons to become tachyonic at the current cosmic density, which leads to the bound $\lambda\gtrsim10^{-96}$ by Relation \eqref{eq:lambda_constraint}. Also, solar system tests imply that the dilaton's coupling constant should be $a_2\gtrsim10^6$~\cite{Brax_2017}, and the potential factor must be $V_0\lesssim10^{-120}\,M_P^4$ to be compatible with the current vacuum energy density~\cite{Khoury_2004a}. However, achieving such small values is computationally costly. 

In our shooting method, we successfully achieved convergence in most cases with a tolerance of $\phi_{\text{tol}}=10^{-10}$, although we relaxed it to $\phi_{\text{tol}}=10^{-7}$ or $\phi_{\text{tol}}=10^{-5}$ in a few cases. Reaching this level of precision required working with 20 decimal places, which is twice the precision of the most restrictive tolerance considered. This was necessary even for parameters such as $\mu = 10^{-40}\,M_P$ and $V_0=10^{-85}\,M_P^4$. Nevertheless, as discussed later, our key conclusions remain valid for parameter values that further suppress the impact of the scalar field.

{In Table~\ref{tab1}, we show the characteristic functions of the screening mechanisms studied in this work and~\cite{bachsesteban2024}, as well as their coupling strengths in effective potential minima, to highlight their differences and affinities. The similarity between the symmetron and the dilaton couplings is evident. The difference in the coupling strength of these fields arises from the very different forms of their potentials. For the symmetron, the decoupling that occurs when the environment density exceeds the critical value, $\rho>\mu^2M_S^2$, becomes obvious since $\beta(\Bar{\phi})=0$ corresponds to that case. When the density is below this threshold, the coupling strength takes the non-zero constant value in the table. The dilaton coupling, on the other hand, does not experience any abrupt changes, as it varies smoothly with the density. Finally, the chameleon's coupling strength is constant, clearly distinguishing it from the other two mechanisms.}

\begin{table}[H] 
\caption{{{The} 
 {three} 
 scalar field models whose behaviour around WDs we have studied between the present paper and~\cite{bachsesteban2024}. We include their coupling functions $A(\phi)$, self-interacting potentials $V(\phi)$, and coupling strengths evaluated at an effective potential minimum $\beta(\Bar{\phi})$.
\textit{{Chameleon}}: Its typical inverse power-law potential is determined by an integer $n\geq1$ and the energy scale $\Lambda$. The dimensionless constant $\beta$ sets the coupling between this scalar field and matter. 
\textit{{Symmetron}}: The potential mass scale $\mu$ and the dimensionless self-coupling parameter $\lambda$ define the symmetry-breaking potential of this mechanism, as well as the value that the field takes at infinity (see Section~\ref{subsec:Symm}). The cutoff scale $M_S$ controls the coupling to matter and, together with the other two parameters, determines its strength. 
\textit{{Dilaton}}: The dimensionless constant $a_2$ controls the dilaton coupling, whose function has a minimum at $\phi_d$. When evaluated at a minimum of the effective potential (see Equation~\eqref{eq:dil_effective_minimum}), the coupling strength depends only on the constant energy density $V_0$ and is inversely proportional to the density.}\label{tab1}}
\setlength{\cellWidtha}{\textwidth/4-2\tabcolsep-0in}
\setlength{\cellWidthb}{\textwidth/4-2\tabcolsep-0in}
\setlength{\cellWidthc}{\textwidth/4-2\tabcolsep-0in}
\setlength{\cellWidthd}{\textwidth/4-2\tabcolsep-0in}

\begin{tabularx}{\textwidth}{
>{\centering\arraybackslash}m{\cellWidtha}
>{\centering\arraybackslash}m{\cellWidthb}
>{\centering\arraybackslash}m{\cellWidthc}
>{\centering\arraybackslash}m{\cellWidthd}
}
\toprule
\textbf{Scalar Field}	& \boldmath{$A(\phi)$}	& \boldmath{$V(\phi)$} & \boldmath{$\beta(\Bar{\phi})$} \\
\midrule
Chameleon		& $e^{\beta\phi/M_P}$    & $\Lambda^4\left(\frac{\Lambda}{\phi}\right)^n$    & $\beta$  \\
Symmetron		& $1+\frac{1}{2}\frac{\phi^2}{M_S^2}
$			& $-\frac{1}{2}\mu^2\phi^2+\frac{1}{4}\lambda\phi^4$ & $0\text{ or }\frac{M_P}{M_S^2}\frac{\mu}{\sqrt{\lambda}}$ \\
Dilaton		& $1+\frac{a_2}{2}\frac{\left(\phi-\phi_d\right)^2}{M_P^2}
$	& $A^4(\phi)V_0e^{-(\phi-\phi_d)/M_P}$ & $\frac{V_0}{4V_0+\rho}$ \\
\bottomrule
\end{tabularx}
\end{table}

 \subsection{\label{subsec:Pressure}Pressure Profiles}
 
\begin{figure}[H]
\begin{tabular}{c}
\includegraphics[width=0.7\linewidth]{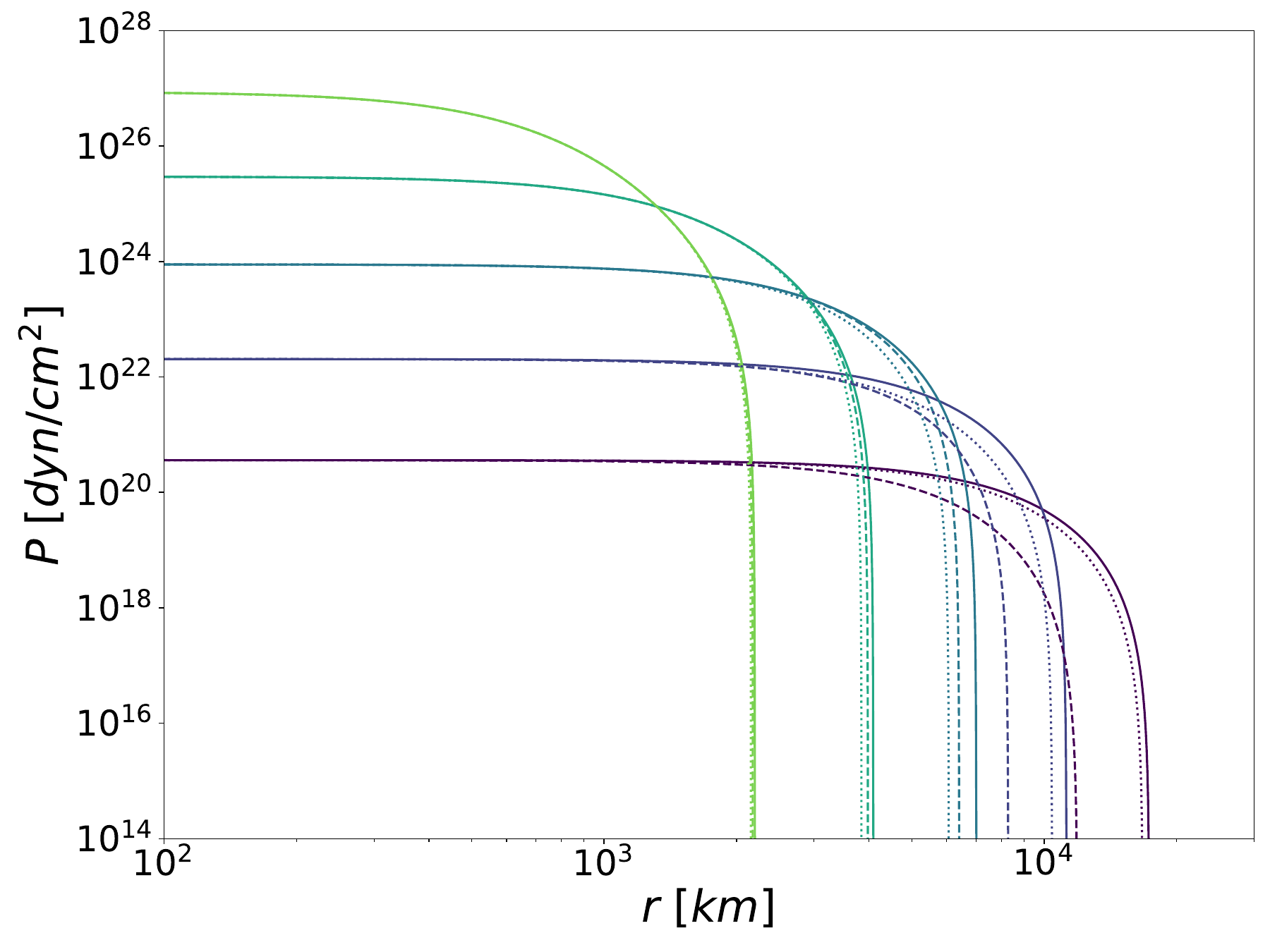}\\
\includegraphics[width=0.7\linewidth]{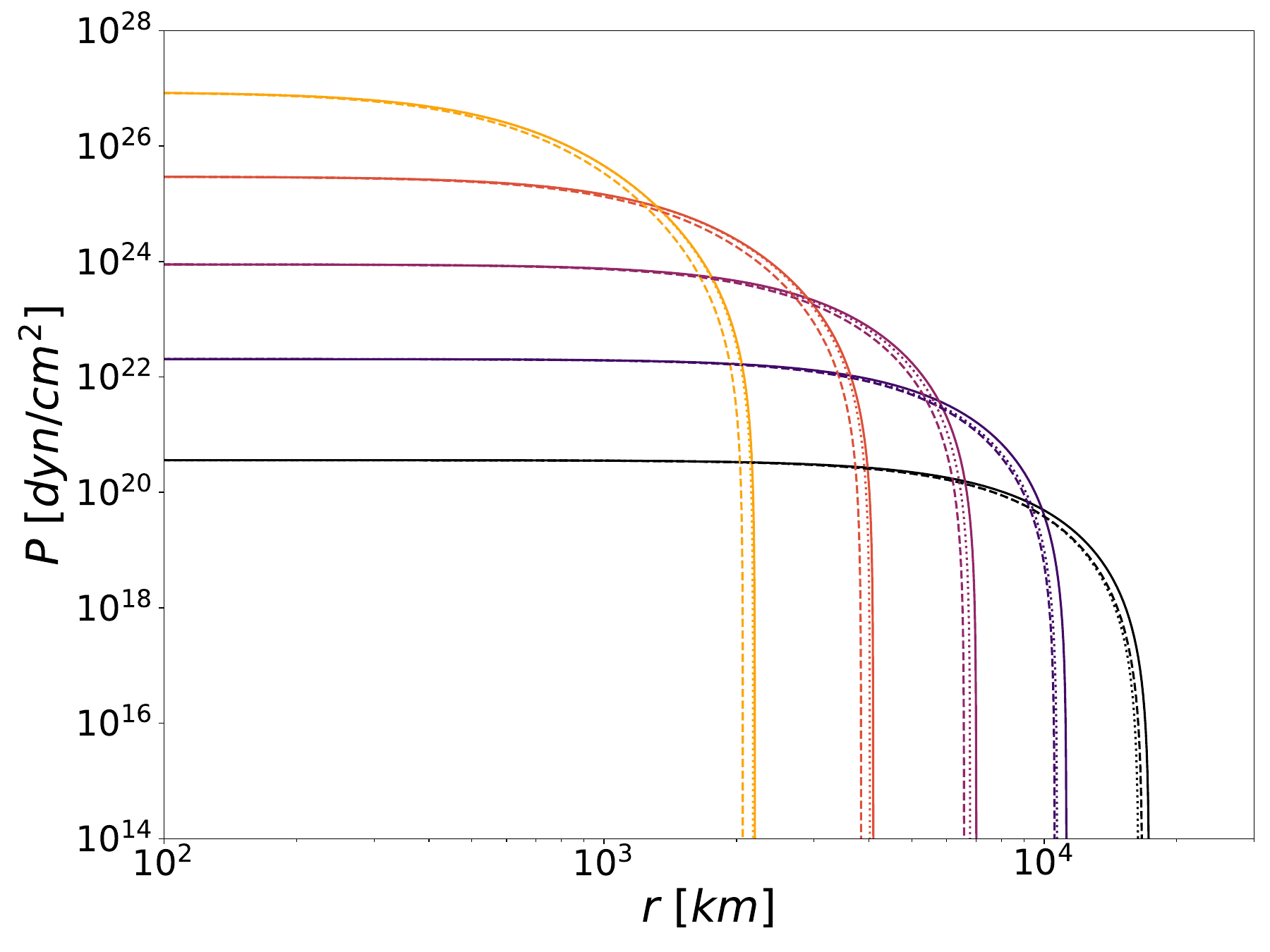}
\end{tabular}
     
    \caption{Radial pressure profiles of WDs screened by the symmetron (\textbf{top}) and dilaton (\textbf{bottom}) mechanisms. For the symmetrons, defined by Equation~\eqref{eq:sym}, we consider $M_S=10^{-2}\,M_P$, \linebreak $\mu=1\times10^{-41}\,M_P$ (dashed) and $\mu=5\times10^{-41}\,M_P$ (dotted), with the corresponding $\lambda$ values given by Equation~\eqref{eq:lambda_constraint}. The dilaton model is characterised by Equations~\eqref{eq:dil_coupling} and \eqref{eq:dil_potential}, with parameters $\phi_d=0$, $a_2=10^1$, $V_0=10^{-85}\,M_P^4$ (dashed) and $a_2=10^3$, $V_0=10^{-87}\,M_P^4$ (dotted). For comparison, WDs in Newtonian gravity results are shown in both panels (solid). The colour brightness corresponds to increasing central densities, ranging from purple to green for symmetrons and from black to yellow for dilatons (these colour schemes are maintained throughout the paper for consistency). Profiles are shown for central densities $\Tilde{\rho}_0 = 7.0\times10^{4},\,8.5\times10^{5},\,1.0\times10^{7},\,1.3\times10^{8},\,1.5\times10^{9}\text{ g}\,\text{cm}^{-3}$.}
    \label{fig:Pvsr}
\end{figure}

The top panel of Figure~\ref{fig:Pvsr} shows the radial pressure profiles $\Tilde{P}(r)$ for WDs in Newtonian gravity for two different realisations of the symmetron model characterised by a different potential mass scale, namely $\mu=1\times10^{-41}\,M_P$ and $\mu=5\times10^{-41}\,M_P$. We fixed the cutoff at $M_S=10^{-2}\,M_P$ and the coupling $\lambda$ is set by Relation \eqref{eq:lambda_constraint}. We observe almost no difference for the densest stars, whereas the departure from Newtonian behaviour depends on $\mu$ for lower densities.
On the bottom panel, we find the same profiles for the dilaton model with $\phi_d=0$, $a_2=10^1$, $V_0=10^{-85}\,M_P^4$ and $a_2=10^3$, $V_0=10^{-87}\,M_P^4$. In the first case, we see a comparable shift in the pressure drop for all central densities. In contrast, in the second model, the pressure drop varies with the star density, moving from the Newtonian scenario for higher densities to the first dilaton case for lower ones.
 
 From the hydrostatic Equation \eqref{eq:background_p_Newton}, we know that the pressure decrease depends on the coupling and the gradient of the scalar field. From Equations~\eqref{eq:sym} and \eqref{eq:dil_coupling}, we know that the coupling functions $A(\phi)$ are always positive for both fields and that their derivatives $A_{,\phi}$ will be so whenever the field $\phi$ is positive. The scalar field reaches a non-negative minimum both inside and outside the star, with the external minimum being higher than the internal one due to the lower density outside the star compared to the inside (for symmetron fields, the internal minimum might be $\Bar{\phi} = 0$; see \mbox{Sections~\ref{subsec:Symm} and \ref{subsec:Dila})}. Because there are no additional extrema between the two minima, the scalar field continuously increases and never becomes negative, which also means that its gradient is always positive (see {Figure}~4
 ~for computational evidence). Therefore, the presence of a scalar field always causes the stellar pressure to decline faster.
 
 As we have already hinted, a symmetron field will be set to zero within the densest WDs. This means that it is effectively decoupled from matter, hence the insignificant change on the highest curve of Figure~\ref{fig:Pvsr}. In contrast, for sufficiently low densities, it acquires a positive VEV, becoming coupled to matter and thus affecting the stellar structure. For intermediate densities, a larger mass scale leads to a greater deviation from Newtonian behaviour, as expected. If we closely examine the figure, we notice that a higher potential mass scale causes the pressure profile of less-dense WDs to resemble the Newtonian gravity profile. We will discuss this phenomenon further in Section~\ref{subsec:MassRadius}, after reviewing all \mbox{the results}.


Regarding the dilaton model, the central densities are comparable to the potential scales. In fact, $\Tilde{\rho}_0 \gtrsim V_0$ for most stars, meaning that the scalar field value is mainly governed by the environmental density (see Equation~\eqref{eq:dil_effective_minimum}). This means that the dilaton field is always positive and coupled to matter, altering the stellar pressure regardless of the central density. In that sense, the dilaton mechanism behaves as the chameleon one~\cite{bachsesteban2024}. In contrast, while the shift in pressure profiles remained constant across densities for chameleon fields, it depends on the parameter choice for dilaton fields. From Equation~\eqref{eq:dil_eff_mass_infty}, we can calculate the effective mass of the dilaton and estimate its impact. For the lowest density shown in Figure~\ref{fig:Pvsr}, the dilaton mass is comparable for both parameter choices we have considered, which explains the similar effect on the pressure. Then, as the density increases, the effective mass increases more for the second set of parameters ($a_2=10^3$, $V_0=10^{-87}\,M_P^4$) than for the first ($a_2=10^1$, $V_0=10^{-85}\,M_P^4$). As a result, in the second case, the dilaton effect becomes increasingly suppressed, approaching the Newtonian case in the densest stars.

In summary, we can anticipate from Figure~\ref{fig:Pvsr} that dense symmetron-screened WDs will be identical to their Newtonian gravity counterparts, while the least dense ones will have smaller radii and, consequently, smaller masses. In the dilaton scenario, we forecast that certain parameter configurations will also lead to stars very similar to their Newtonian analogues, particularly along the upper part of the MR curve, but we generally expect lower stellar masses and radii (see Section~\ref{subsec:MassRadius}). As we previously explained for both scalar fields under consideration---as well as for the chameleon mechanism in~\cite{bachsesteban2024}---this kind of scalar field will never produce super-Chandrasekhar WDs.

\subsection{\label{subsec:CoolTime}Cooling Time}

The presence of a scalar field is expected to influence the thermal properties of WDs. To quantify this effect, we calculate the mean specific heat, given by
\begin{equation}\label{eq:specific_heat}
    \Bar{c}_V=\frac{1}{M}\int_0^M\left(c_V^{\text{ion}}+c_V^{\text{el}}\right)dm\,,
\end{equation}
where $M$ is the stellar mass, $c_V^{\text{ions}}$ represents the specific heat of ions, and $c_V^{\text{el}}$ corresponds to that of the electrons~\cite{Kalita_2023}. The specific heat of ions depends on the Coulomb-to-thermal energy ratio, $\Gamma$, with a critical value of $\Gamma_c=125$~\cite{Brush_1966}. If $\Gamma < \Gamma_c$, the specific heat of ions remains constant at $c_V^{\text{ion}}=(3/2)k_B=3/2$. For $\Gamma > \Gamma_c$, it becomes temperature-dependent and is given by
\begin{equation}\label{eq:ion_heat}
    c_V^{\text{ion}}=9\left(\frac{T}{\Theta_D}\right)^3\int_0^{\Theta_D/T}\frac{x^4e^x}{(e^x-1)^2}dx\,,
\end{equation}
where the Debye temperature, $\Theta_D$, is expressed in $K$ as
\begin{equation}\label{eq:Debye_T}
    \Theta_D=3.48\times10^3 Y_e\sqrt{\Tilde{\rho}}
\end{equation}
with the stellar density $\Tilde{\rho}$ in $\text{g}\,\text{cm}^{-3}$. The specific heat of electrons depends on temperature too and is described by~\cite{Koester_1972}
\begin{equation}\label{eq:e_heat}
    c_V^{\text{el}}=\frac{\pi^2}{2}Z\frac{T}{\Tilde{\epsilon}_F}\,,
\end{equation}
where $\Tilde{\epsilon}_F = \Tilde{p}_F^2 + m_e^2$ is the Fermi energy, the Fermi momentum is defined by \linebreak $\Tilde{p}_F^3=3\pi^2Y_e\Tilde{\rho}/m_p$, and $m_p$ is the proton mass (see Section~\ref{subsec:EoS}). 

Consequently, different specific heats lead to different cool-downs. To demonstrate this explicitly, we consider the luminosity $L$ of WDs, given by
\begin{equation}\label{eq:luminosity}
    L=-\frac{M}{Am_p}\Bar{c}_V\frac{dT}{dt}\,,
\end{equation}
where $M$ is the stellar mass, $T$ is the effective temperature of the star, and the luminosity comes primarily from the loss of thermal energy from ions and electrons over time $t$. To solve this equation, an additional relation between the luminosity and the effective temperature is required. This can be derived by integrating the structure equations for a WD. A discussion of this topic, along with various $L-T$ relationships incorporating opacity models and convective effects in the outer layers, is provided in~\cite{Koester_1990}. From that study, the following empirical luminosity fit is obtained:
\begin{equation}\label{eq:L/M}
    \frac{L}{M}=9.743\times10^{-21}T^{2.56}\frac{L_\odot}{M_\odot}\,,
\end{equation}
where the temperature $T$ is given in Kelvin. The dimming of WDs over time is shown in {Figure}~3
, assuming an initial temperature of $T_\text{ini}=10^8\,K$ and allowing cooling down to a final temperature of $T_\text{ini}=10^6\,K$. The time taken between these two points represents the cooling time of the WD.

\begin{figure}[H]
\begin{tabular}{c}
\includegraphics[width=0.8\linewidth]{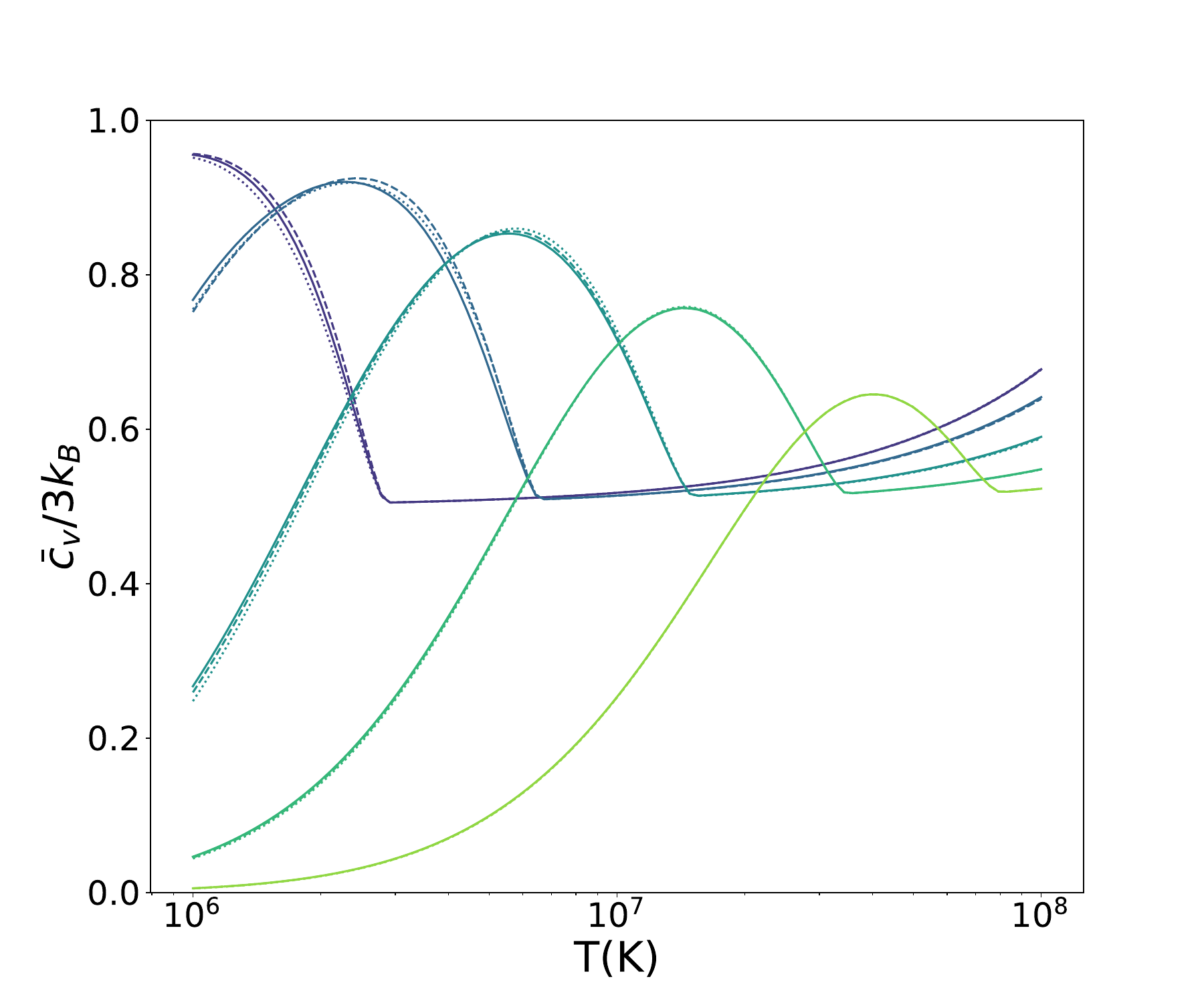}\\
\includegraphics[width=0.8\linewidth]{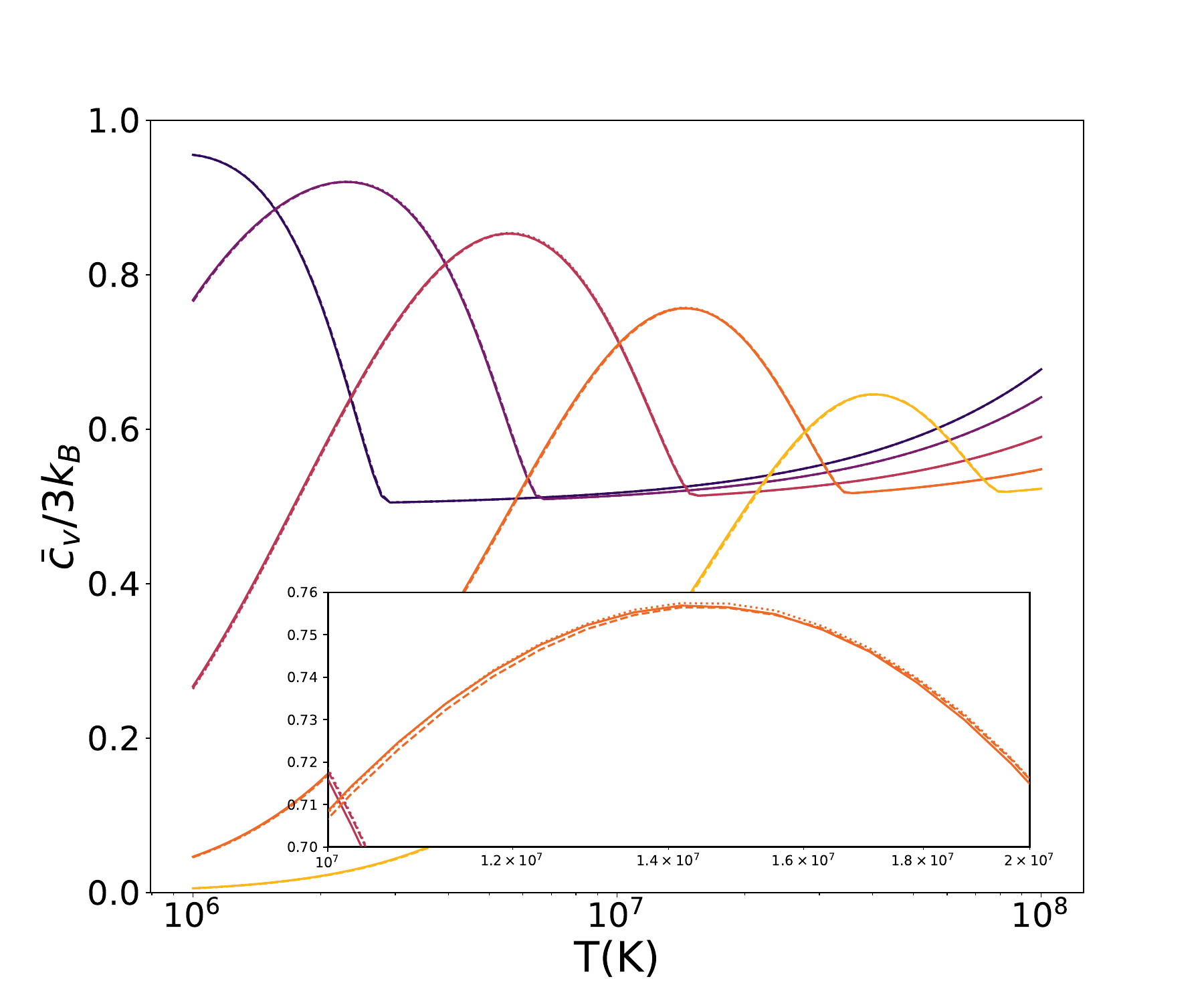}
\end{tabular}
%
    \caption{{Mean} 
 specific heat $\Bar{c}_V$ as a function of temperature $T$ of WDs screened by the symmetron (\textbf{top}) and dilaton (\textbf{bottom}) mechanisms. For the symmetron model, defined by Equation~\eqref{eq:sym}, we consider $M_S=10^{-2}\,M_P$, $\mu=1\times10^{-41}\,M_P$ (dashed) and $\mu=5\times10^{-41}\,M_P$ (dotted), with the corresponding $\lambda$ values given by Equation~\eqref{eq:lambda_constraint}. The dilaton model is characterised by Equations~\eqref{eq:dil_coupling} and \eqref{eq:dil_potential}, with parameters $\phi_d=0$, $a_2=10^2$, $V_0=10^{-85}\,M_P^4$ (dashed) and $a_2=10^3$, $V_0=10^{-87}\,M_P^4$ (dotted). Results for Newtonian gravity WDs are provided for reference (solid). The colour brightness indicates increasing central densities, transitioning from purple to green for symmetrons and from black to yellow for dilatons. Profiles are shown for central densities $\Tilde{\rho}_0$ = $4.5\times10^{5},\,5.5\times10^{6},\,$\linebreak $6.7\times10^{7},\,8.2\times10^{8},\,1.0\times10^{10}\text{ g}\,\text{cm}^{-3}$.}
    \label{fig:CvsT}
\end{figure}

The relationship between temperature and mean specific heat for carbon WDs is shown in Figure~\ref{fig:CvsT} for Newtonian WDs and the same models we explored in Figure~\ref{fig:Pvsr}: $\mu=1\times10^{-41}\,M_P$ and $\mu=5\times10^{-41}\,M_P$ with $M_S=10^{-2}\,M_P$ and the corresponding $\lambda$ (see Equation~\eqref{eq:lambda_constraint}) for the symmetron mechanism (top panel), ($a_2=10^1$, $V_0=10^{-85}\,M_P^4$) and ($a_2=10^3$, $V_0=10^{-87}\,M_P^4$) with $\phi_d=0$ for the dilaton mechanism (bottom panel). Similar to the pressure profiles, there is no noticeable difference between the Newtonian $\Bar{c}_V-T$ curve and those affected by symmetron fields for the densest stars. However, for WDs with lower densities, the $\Bar{c}_V-T$ curves are shifted to higher temperatures and higher specific heat values, indicating that the maximum $\Bar{c}_V$ is both greater than in the Newtonian case and occurs at a higher temperature. This trend does not hold universally, as the $\mu=5\times10^{-41}\,M_P$ curve for the least dense star falls below its Newtonian counterpart. As with the pressure profiles, a larger potential mass scale $\mu$ does not always correspond to a greater deviation from the Newtonian results. This is evident from the switch between the two symmetron realisations for WDs of intermediate densities. On the other hand, there are no significant differences between the WDs affected by the dilaton and those that are not. As we can see from the zoomed-in section of the plot, both dilaton models align with the Newtonian curve.

\begin{figure}[H]
\begin{tabular}{c}
\includegraphics[width=0.69\linewidth]{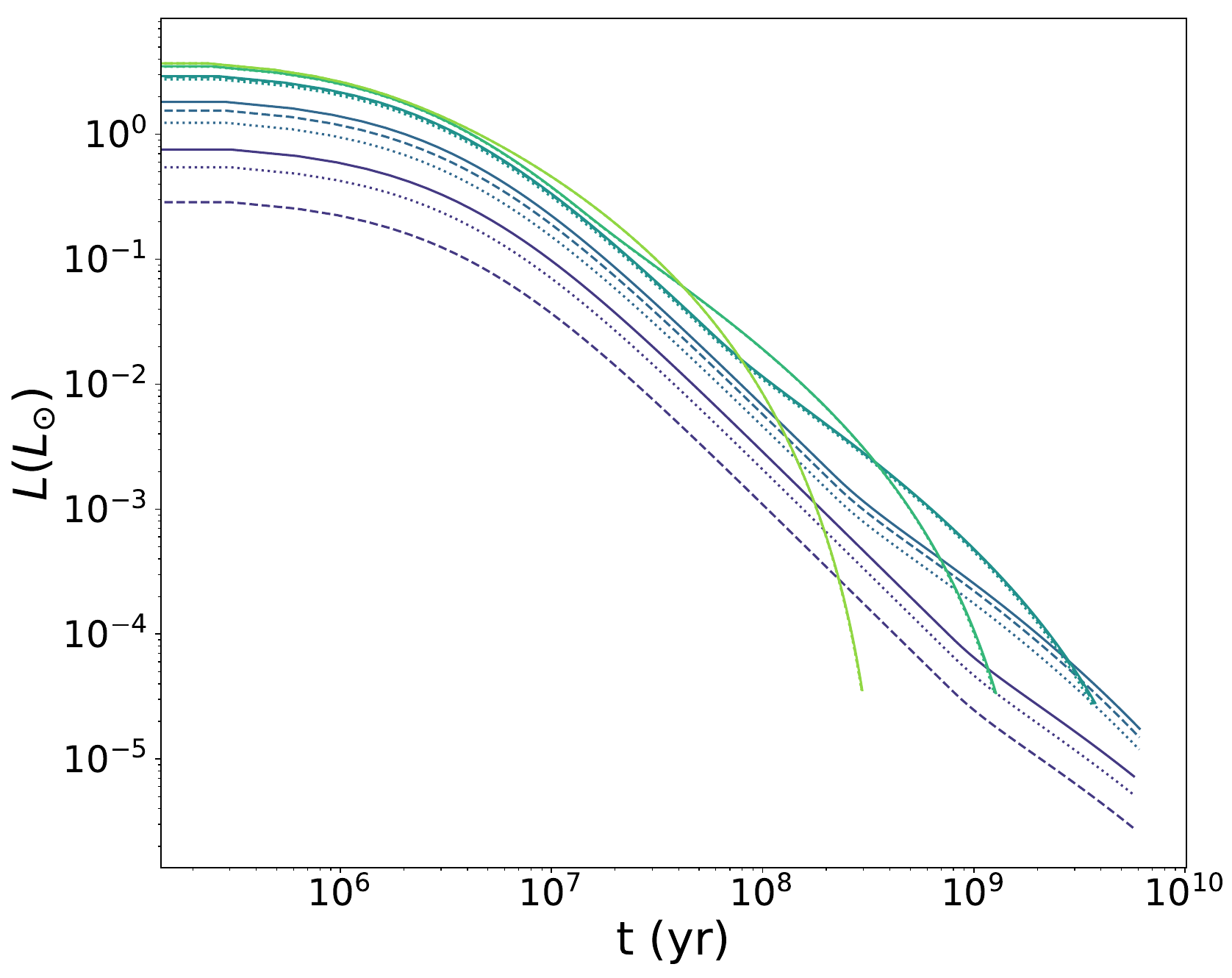}\\
\includegraphics[width=0.69\linewidth]{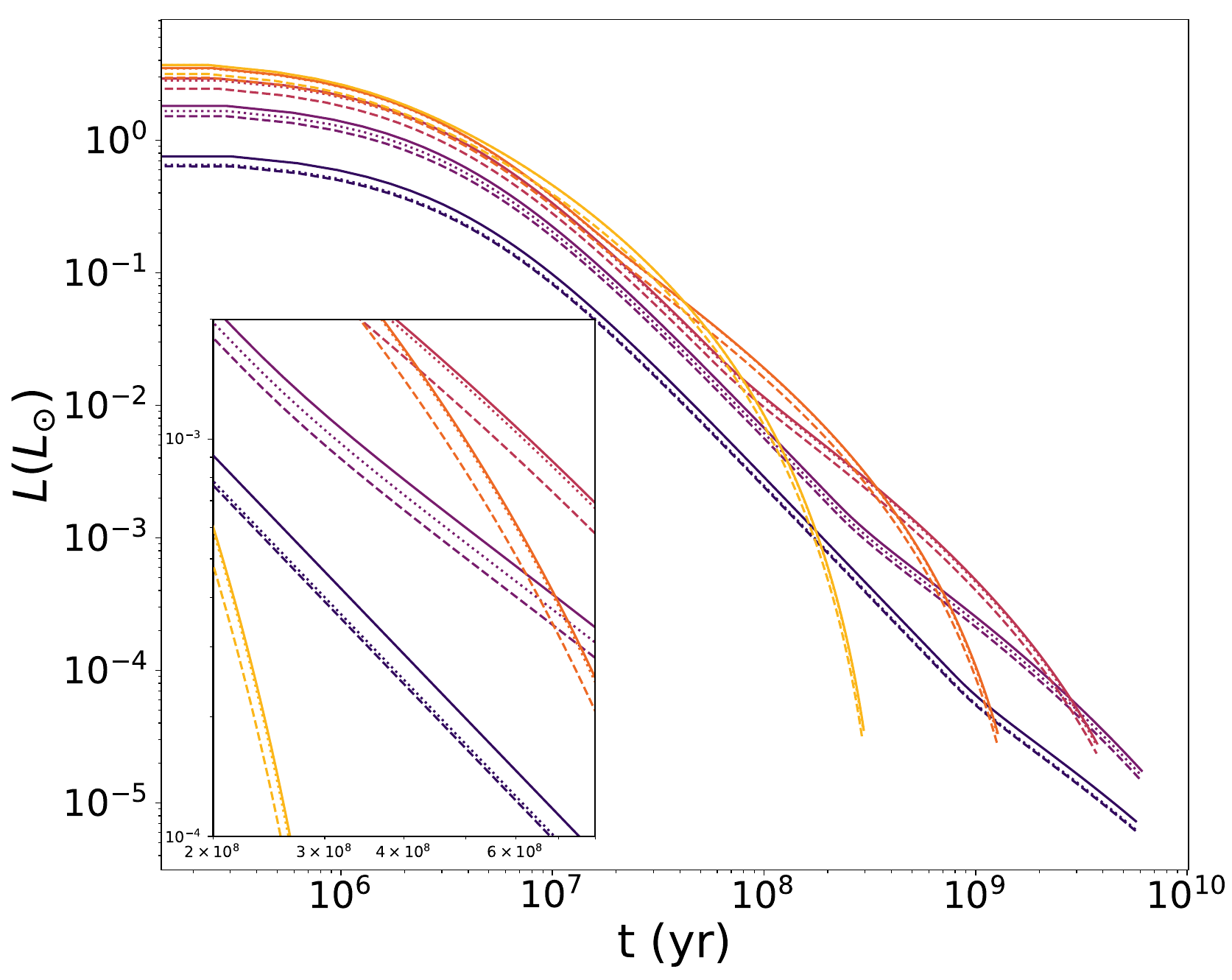}
\end{tabular}
%
    \caption{{Luminosity} 
 $L$ as a function of time $t$ of WDs screened by the symmetron (\textbf{top}) and dilaton (\textbf{bottom}) mechanisms. For symmetron fields (see Equation~\eqref{eq:sym}), we consider $M_S=10^{-2}\,M_P$, $\mu=1\times10^{-41}\,M_P$ (dashed) and $\mu=5\times10^{-41}\,M_P$ (dotted), and the corresponding $\lambda$ couplings given by Equation~\eqref{eq:lambda_constraint}. For dilaton fields (see Equations~\eqref{eq:dil_coupling} and \eqref{eq:dil_potential}), we choose $\phi_d=0$, $a_2=10^2$, $V_0=10^{-85}\,M_P^4$ (dashed) and $a_2=10^3$, $V_0=10^{-87}\,M_P^4$ (dotted). Results for Newtonian WDs are also plotted for reference (solid). Colour brightness indicates increasing central densities, changing from purple to green for symmetrons and from black to yellow for dilatons. Profiles are shown for central densities $\Tilde{\rho}_0 = 4.5\times10^{5},\,5.5\times10^{6},\,6.7\times10^{7},\,8.2\times10^{8},\,1.0\times10^{10}\text{ g}\,\text{cm}^{-3}$}
    \label{fig:Lvst}
\end{figure}

In Figure~\ref{fig:Lvst}, we display the luminosity decay over time for the same scenarios studied in Figures~\ref{fig:Pvsr} and~\ref{fig:CvsT}. Regarding the symmetron scenario (top panel), we observe the expected behaviour based on the specific heat curves. For the highest densities, there are no significant differences between WDs with a scalar field effect and those without. As the density decreases, symmetrons shift the luminosity curves downward. However, we see that they maintain a constant distance from the curves of Newtonian stars. Since we expect smaller and less massive stars---as discussed in Section~\ref{subsec:Pressure}---and luminosity is directly proportional to stellar mass (see Equation~\eqref{eq:luminosity}), they have a lower luminosity from the outset, shifting the entire curve downward. Once again, although the curves of the symmetron-screened WDs lie below the purely Newtonian ones, a higher potential mass scale $\mu$ does not always imply a greater deviation from the Newtonian case.

Concerning the dilaton mechanism (bottom panel), differences from the Newtonian case are observed in terms of luminosity, unlike with mean specific heat. This can also be explained thanks to the effect of the stellar mass, as discussed in the previous paragraph, as dilaton-screened WDs will always be the same size as or smaller than those without a scalar component. As was the case with the pressure curves, we can see in the zoomed-in area that the luminosity curves for the ($a_2=10^1$, $V_0=10^{-85}\,M_P^4$) model remain separated from the Newtonian curves for all densities, whereas those for the ($a_2=10^3$, $V_0=10^{-87}\,M_P^4$) model transition between the Newtonian case and the first model as density decreases. As argued in Section~\ref{subsec:Pressure}, this is due to the change in effective mass.

\subsection{\label{subsec:ScalarProfi}Scalar Profiles}

\begin{figure}[H]
\begin{tabular}{cc}
\includegraphics[width=0.47\linewidth]{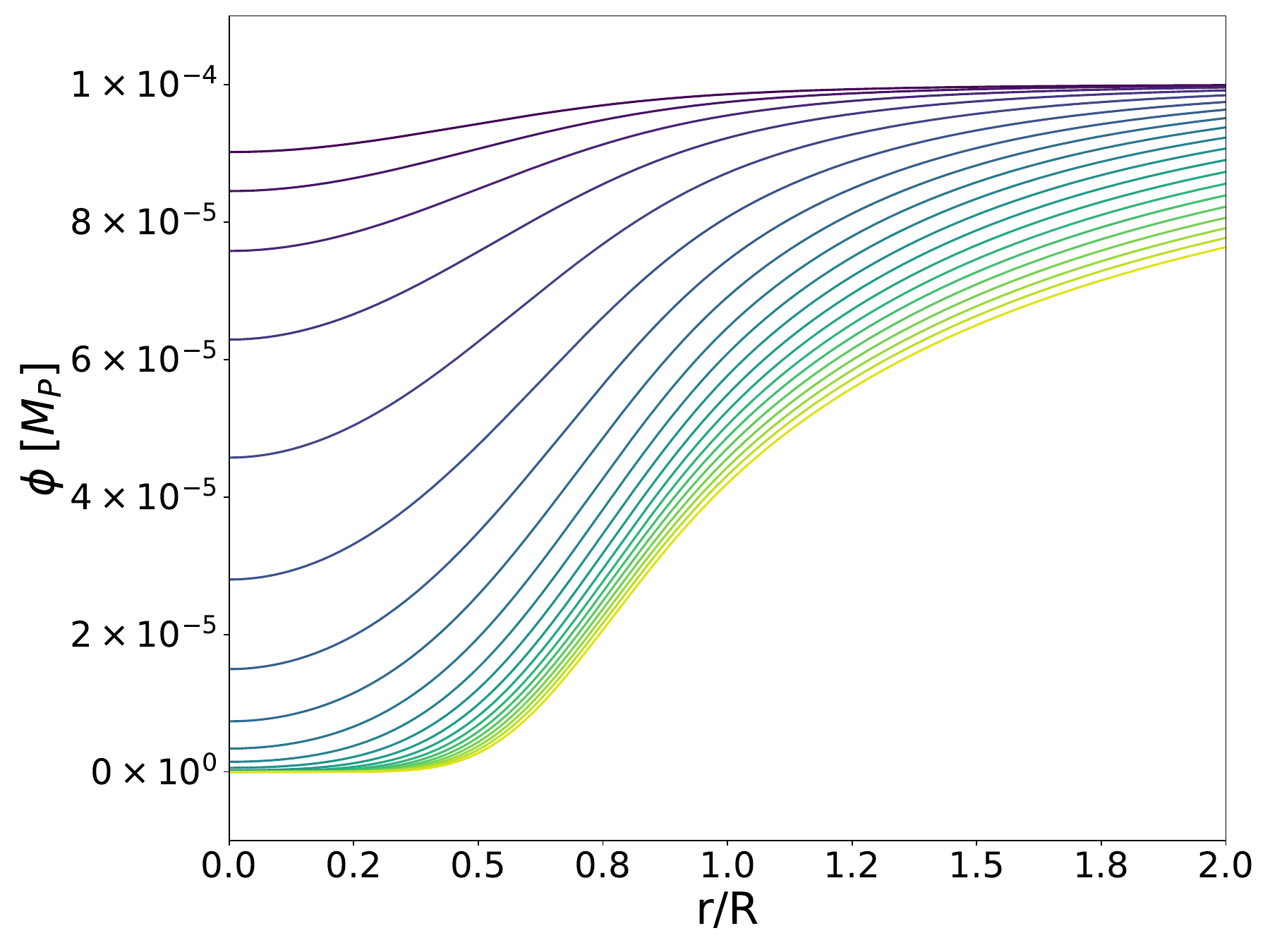}
&\includegraphics[width=0.47\linewidth]{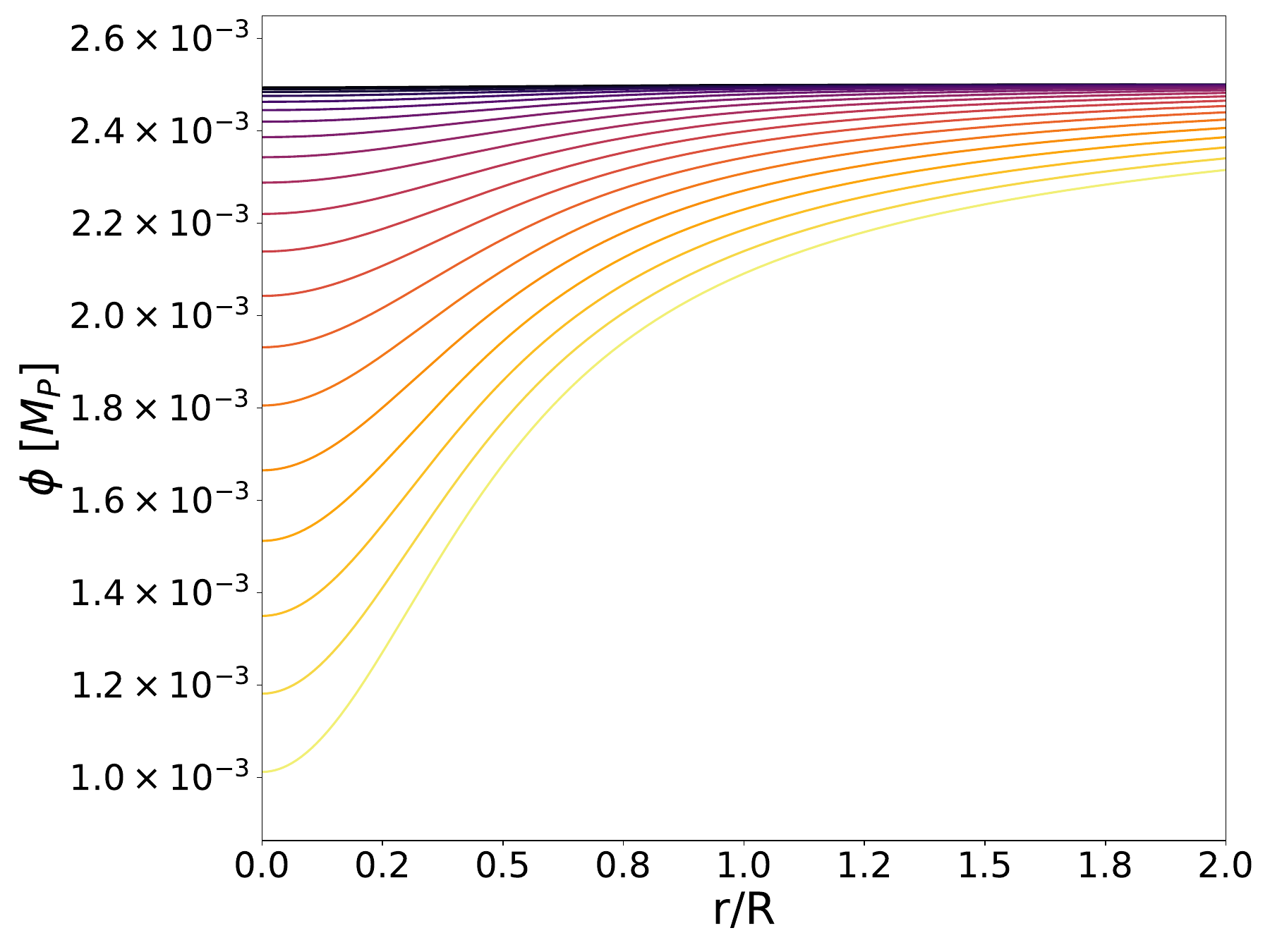}\\

\includegraphics[width=0.47\linewidth]{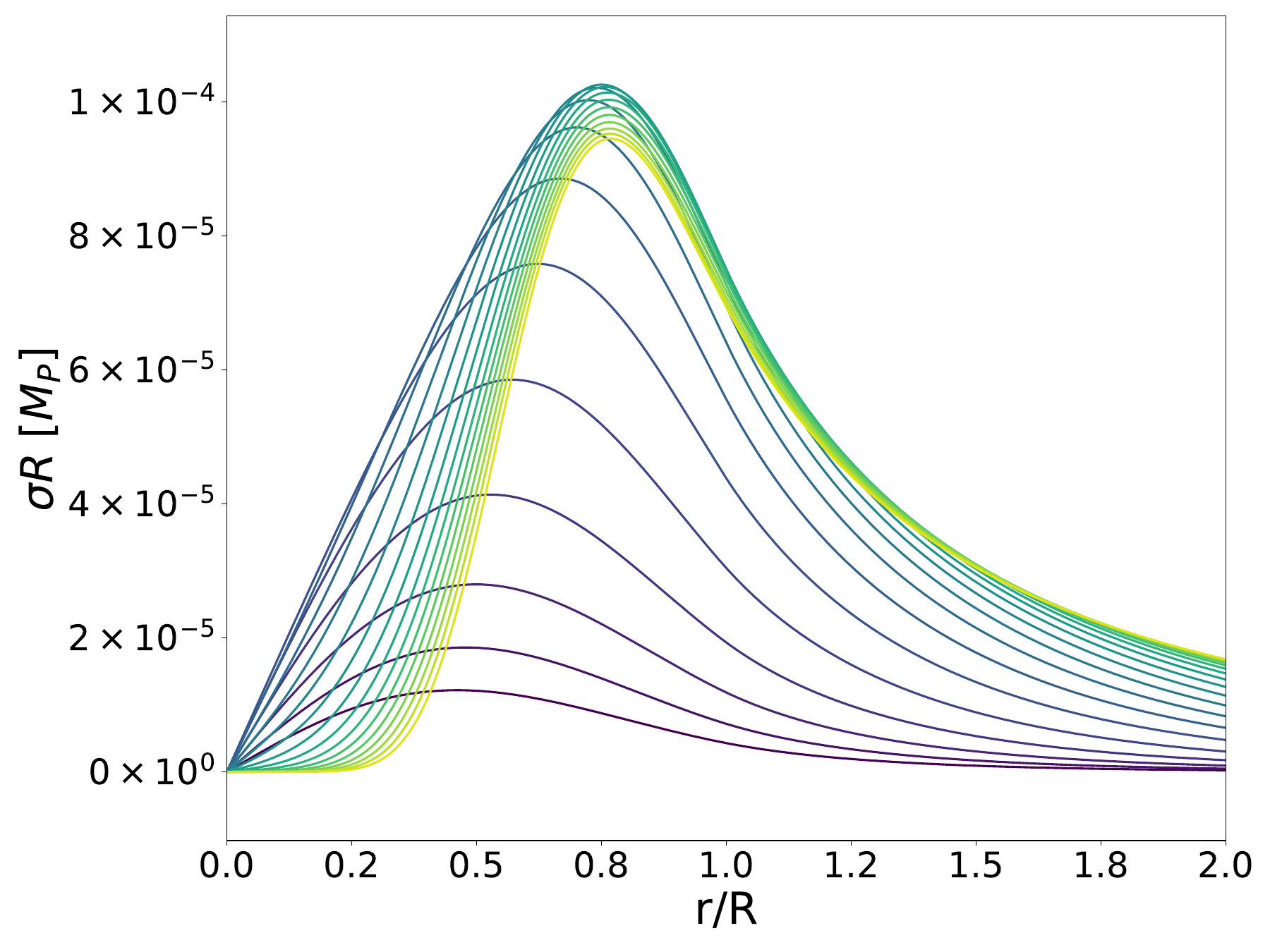}
&\includegraphics[width=0.47\linewidth]{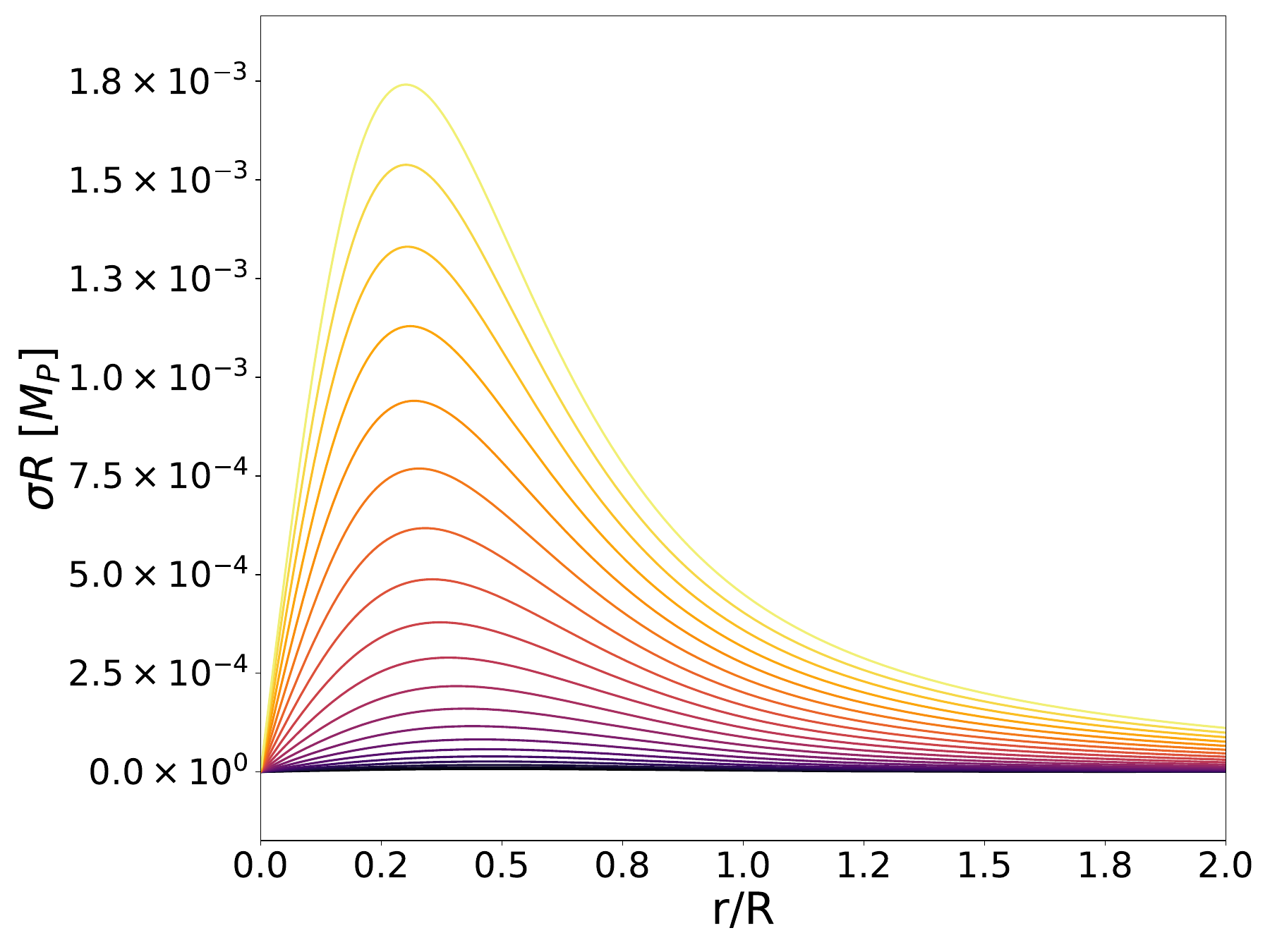}\\

\end{tabular}

    \caption{{Scalar} 
 field (\textbf{top}) and scalar field gradient (\textbf{bottom}) profiles for the symmetron (\textbf{left}) and dilaton (\textbf{right}) mechanisms. The radial coordinate is normalised by the stellar radius $R$ of the respective star, while the gradient is scaled by the same factor. We consider the symmetron model \eqref{eq:sym} with $M_S=10^{-2}\,M_P$, $\mu=1\times10^{-41}\,M_P$, and the $\lambda$ given by Equation~\eqref{eq:lambda_constraint}, and the dilaton model \eqref{eq:dil_coupling} and \eqref{eq:dil_potential} with $\phi_d=0$, $a_2=10^2$, and $V_0=10^{-85}\,M_P^4$. Tones from purple to green (symmetrons) and from black to yellow (dilatons) indicate increasing central densities from $\Tilde{\rho}_{0,\text{min}} = 7\times10^{4} \text{ g}\,\text{cm}^{-3}$ to $\Tilde{\rho}_{0,\text{max}} = 10^{10} \text{ g}\,\text{cm}^{-3}$.}
    \label{fig:phi&sigmavsrR}
\end{figure}

We display radial profiles for the scalar field $\phi(r)$ and its gradient $\sigma(r)$ for symmetron and dilaton fields in Figure~\ref{fig:phi&sigmavsrR}. The symmetron model is fixed by Equation~\eqref{eq:sym} and the parameter choices $M_S=10^{-2}\,M_P$, $\mu=1\times10^{-41}\,M_P$, and the $\lambda$ determined by Expression~\eqref{eq:lambda_constraint}. The dilaton model is given by Equations~\eqref{eq:dil_coupling} and \eqref{eq:dil_potential} with the parameters $\phi_d=0$, $a_2=10^2$, and $V_0=10^{-85}\,M_P^4$. The scalar field profile is approximately flat for low central densities (darker colours) in both cases. As the central density increases, the scalar field becomes suppressed inside the star, particularly in the symmetron mechanisms.

\textls[-20]{As predicted by theory (recall Section~\ref{subsec:Symm}) and anticipated from the previously discussed results, the symmetron field is zero inside the densest stars, where it remains at this value up to half the stellar radius. This means that only the outer shell of the star is affected by the scalar field, while the field itself is screened in the interior. This also translates to the field gradient, which remains consistently zero throughout much of the interior of the densest WDs. }


Regarding the dilaton case, its value at the core of the star depends on the density of the latter, decreasing as the density rises but never reaching zero. We observe that the profiles presented here are very similar to those obtained in the chameleon scenario in~\cite{bachsesteban2024}, which is expected since, as seen and discussed with the pressure profiles in Section~\ref{subsec:Pressure}, the dilaton and chameleon mechanisms exhibit similar behaviour. This can be explained by the fact that both are built from a runaway potential, as explained in Section~\ref{subsec:Dila}.

\subsection{\label{subsec:MassRadius}Mass--Radius Relation}

\begin{figure}[H]
\begin{tabular}{l}
\includegraphics[width=0.65\linewidth]{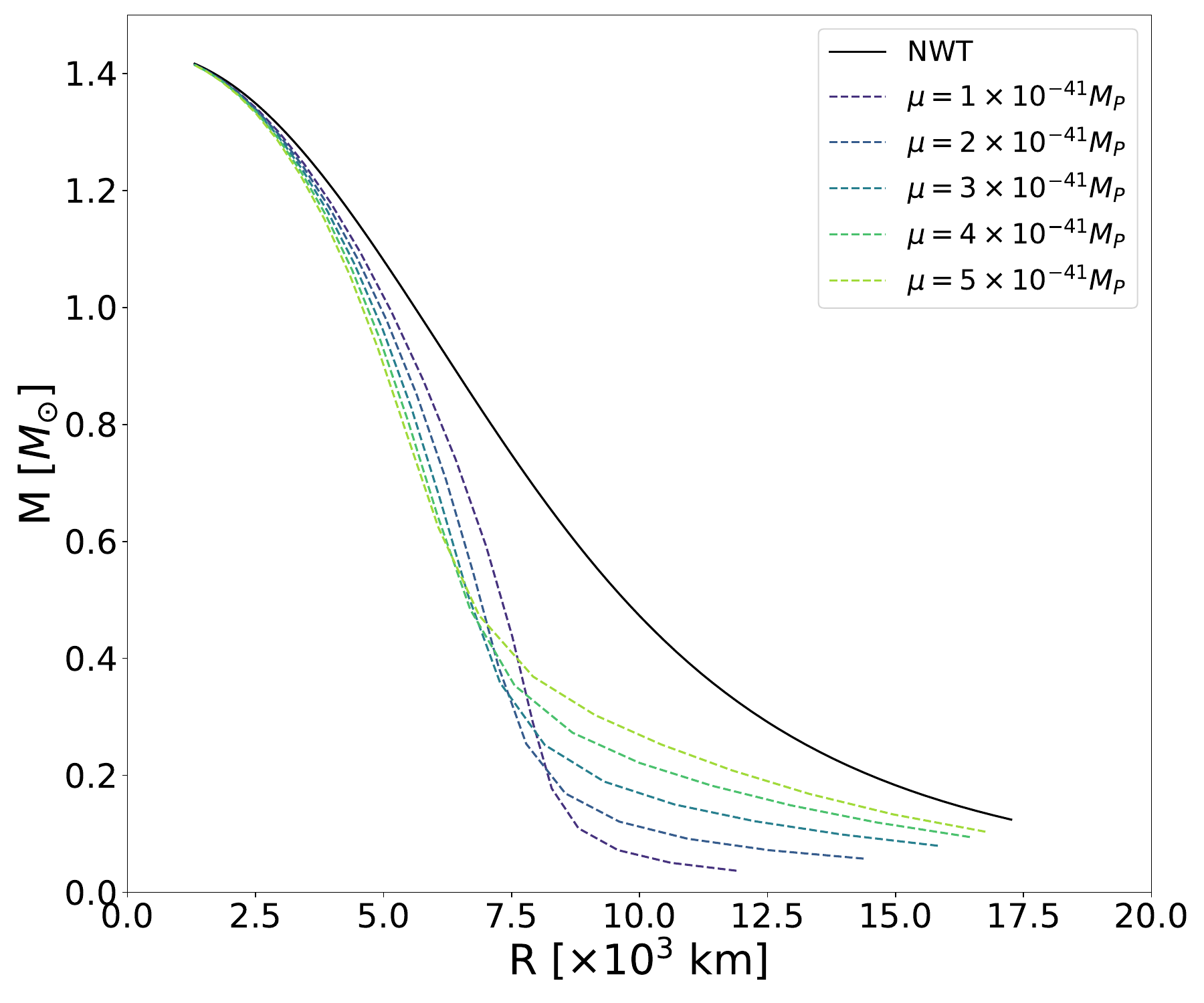}\\
\includegraphics[width=0.65\linewidth]{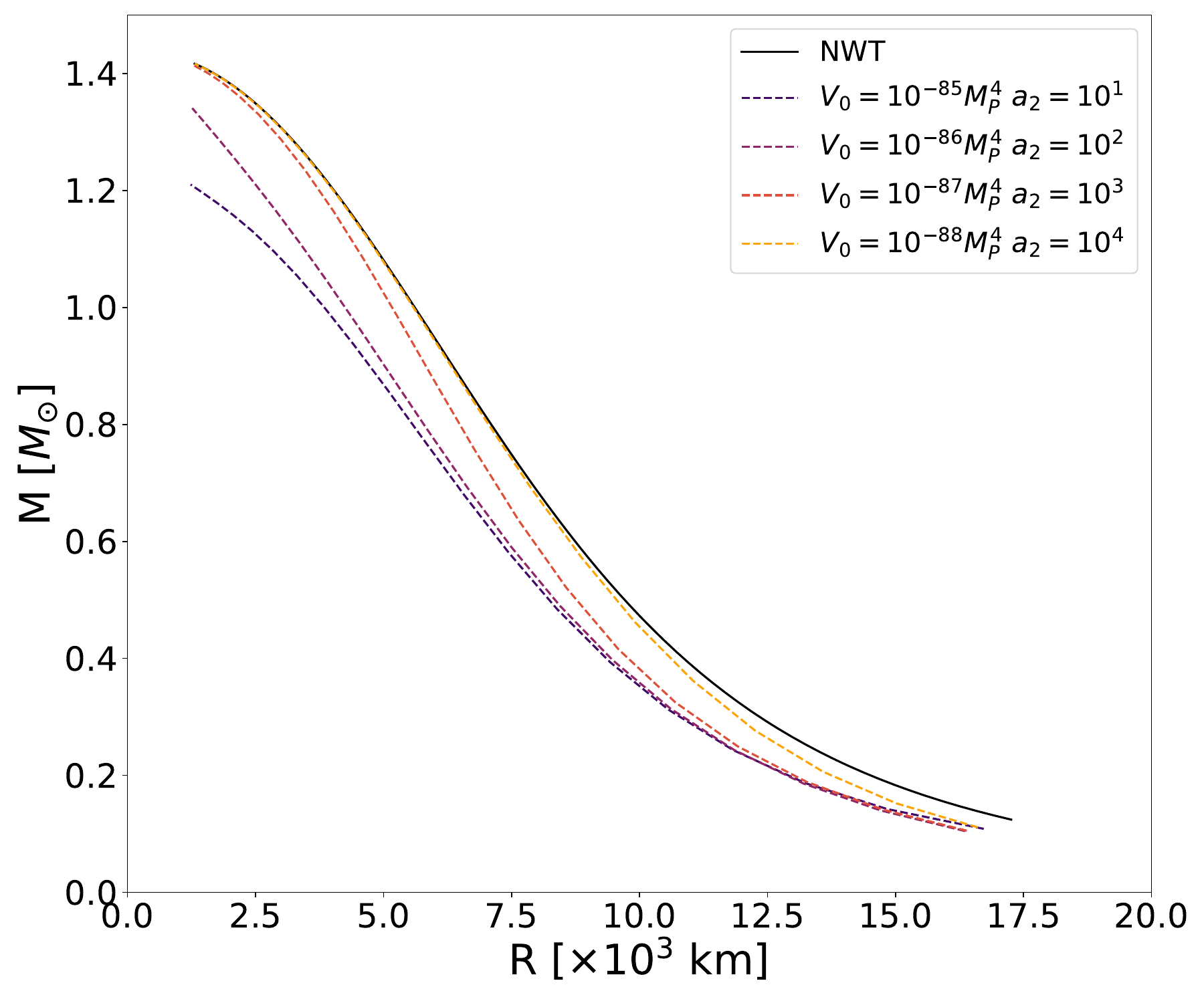}
\end{tabular}
    \caption{\textls[-15]{\textbf{Top panel}: MR curves for symmetron-screened WDs with $M_S=10^{-2}\,M_P$, $\mu=1-5\times10^{-41}\,M_P$, and the corresponding $\lambda$ values determined by Equation~\eqref{eq:lambda_constraint}. \textbf{Bottom panel}: MR curves for dilaton-screened WDs with $\phi_d=0$, $a_2=10^2$, and $V_0=1-10\times10^{-85}\,M_P^4$. In both panels, brighter colours indicate larger potential values, from purple to green for symmetrons and from black to yellow for dilatons. We display in black the MR curve for WDs in Newtonian gravity, for reference.}}\label{fig:MR_curves}
\end{figure}

In Figure~\ref{fig:MR_curves}, we show numerically computed MR curves for symmetron- and dilaton-screened WDs with different parameter choices, within the ranges we discussed at the beginning of the present section. As we inferred from the pressure profiles in Figure~\ref{fig:Pvsr} and reiterated when talking about the luminosity curves in Figure~\ref{fig:Lvst}, no MR curve of symmetron- or dilaton-screened WDs can exceed the one predicted by Newtonian gravity, as is also the case in the chameleon scenario~\cite{bachsesteban2024}. Still, while the MR curves for chameleon-screened WDs followed the shape of the Newtonian one and exhibited  asymptotic behaviour---never crossing nor lying on top of it, regardless of how small the coupling strength and energy scale were---this is not the case for symmetron or dilaton fields.

For the symmetron model (top panel of Figure~\ref{fig:MR_curves}), all MR curves coincide with the Newtonian curve at the massive end. As explained in Section~\ref{subsec:Symm} and shown in Figure~\ref{fig:phi&sigmavsrR}, when the density exceeds the critical value $\mu^2M_S^2$, the symmetron field vanishes and decouples from matter. This explains why there is no significant effect for the densest stars, regardless of the potential mass scale $\mu$. 

As the density decreases and the symmetron field becomes active, the MR curves diverge from the Newtonian one and each other. Initially, the deviation from Newtonian behaviour is proportional to $\mu$. Yet, as observed in the other results contained in this section, this trend reverses for the least massive stars. In lower-density WDs, the density contrast between the interior and exterior is smaller, leading to reduced variation in the field across the WD and beyond (recall Figure~\ref{fig:phi&sigmavsrR}). Consequently, the scalar field gradient---which directly affects the stellar structure (see Equation~\eqref{eq:background_p_Newton})---is also lower. This is why the MR curves tend to approach the Newtonian one again at the lowest densities.

For WDs with intermediate densities, the system lies between these two extremes: symmetron fields are not completely screened in the stellar interior, yet the contrast remains large enough for them to have a significant effect. Nevertheless, it remains to be explained why MR curves get closer to the Newtonian one for higher values of $\mu$ at the less massive end. Looking again at the top panel of Figure~\ref{fig:MR_curves}, we observe that the MR curves for different symmetron realisations intersect. This degeneracy is not present in the chameleon or dilaton cases, where the curves may be very close to each other but never actually cross. From Equation~\eqref{eq:sym_eff_masses}, we see that the symmetron effective mass is directly proportional to $\mu$ when the density is below $\mu^2M_S^2$. Thus, in the lower part of the curve, the field is more suppressed as the potential mass scale increases. This explains why, for instance, the tail of the $\mu=5\times10^{-41}\,M_P$ almost reaches the Newtonian one.

Turning to the dilaton mechanism, we discussed in Sections~\ref{subsec:Pressure} and~\ref{subsec:CoolTime} that the difference between the Newtonian and the dilaton scenarios can be minimal---if not negligible (see Figure~\ref{fig:CvsT})---depending on the choice of the parameters, especially for the densest stars. This explains why the MR curve corresponding to the dilaton model ($a_2=10^4$, $V_0=10^{-88}\,M_P^4$) in the bottom panel of Figure~\ref{fig:MR_curves} almost coincides with the Newtonian one. The greatest overlap is found in the upper half of the curve, where the most massive and densest stars are located. We also see that the upper end of the ($a_2=10^3$, $V_0=10^{-87}\,M_P^4$) curve aligns with the Newtonian case.

Particularly,  we explained in Section~\ref{subsec:Pressure} that the effective mass of the dilaton field increases with density more rapidly for the ($a_2=10^3$, $V_0=10^{-87}\,M_P^4$) model than for ($a_2=10^1$, $V_0=10^{-85}\,M_P^4$). As we move from one model to another following the descending order in the legend, the growth of the effective mass with density becomes more pronounced. This explains why the curve corresponding to the first model ($a_2=10^1$, $V_0=10^{-85}\,M_P^4$) deviates the most from the Newtonian case and why this separation is greater at the massive end of the curve. In this scenario, the dilaton effective mass is significantly smaller than in the other realisations---up to 30 times smaller compared to the fourth parameter set---resulting in weaker field suppression.

\section{\label{sec:Conclusion}Conclusions}

In this study, we investigated the impact of symmetron and dilaton screening on the structure of WDs. The latter are promising candidates for testing alternative gravity theories due to the wealth of available observational data from \textit{{Gaia}} and our well-established understanding of the EoS governing their internal matter. Using a Chandrasekhar EoS, we solved the equilibrium equations using a custom-designed shooting method. Throughout the paper, we have compared the results for these two fields with those obtained for the chameleon in our previous work~\cite{bachsesteban2024}.

We have seen that symmetron and dilaton fields affect the pressure of the least dense WDs in all the configurations we have explored, causing it to decrease more rapidly. This effect is suppressed for the more massive star, even disappearing entirely for symmetrons and certain dilatons. In the case of the symmetron mechanism, this occurs because the field becomes zero and decouples from matter beyond a certain density ($\rho>\mu^2M_S^2$). The dilaton field does not vanish completely, but the small values it adopts lead to a very weak coupling strength (see the expression after Equation~\eqref{eq:dil_coupling}). As we mentioned in Section~\ref{sec:intro} and explained in Sections~\ref{subsec:Symm} and~\ref{subsec:Dila}, this is a common feature of both fields: they are screening mechanisms characterised by an environment-dependent coupling. In cases where the scalar field interacts negligibly with matter, the results converge to the Newtonian ones. This means that the pressure decreases later than in the Newtonian case in none of the studied scenarios.

The fact that the pressure always decreases at the same rate or earlier than in the Newtonian situation implies that none of these screening mechanisms---neither the symmetron, the dilaton, nor the chameleon---can produce stars that are larger or more massive than those predicted by Newtonian theory. As seen in Figure~\ref{fig:MR_curves} and previously in~\cite{bachsesteban2024}, all MR curves for stars with a scalar field contribution lie below the Newtonian one. Nevertheless, these curves have allowed us to observe how different parameter choices can lead to seemingly counterintuitive behaviours. Regarding the symmetron model, a higher potential mass scale $\mu$ results in a smaller deviation from the Newtonian case for the smallest WDs. For dilaton models, we have found that, when the field is not screened at high densities, the MR curve deviates more from the Newtonian case for massive stars than for smaller ones---a trend also observed for chameleon models, although less pronounced.

As highlighted throughout this paper, the dilaton and chameleon mechanisms share similarities, as both are built upon runaway potentials. Consequently, we have confirmed that the radial profiles of the dilaton model and its gradient resemble those obtained for the chameleon one. In both cases, the field value at the centre of the star depends on its density. In contrast, for the symmetron mechanism, the field settles to zero beyond the aforementioned critical density, at the star centre and throughout much of its core.

Regarding luminosity, we have found that it is also lower than or equal to that predicted in the Newtonian case, following the density dependence we have explained here concerning changes in pressure and mass, and as discussed throughout this work. We have verified that these changes are primarily due to the lower stellar masses obtained in the case of symmetron- and dilaton-screened WDs, as the specific heat showed few significant differences, particularly for the dilaton models we considered.

This study has allowed us to identify phenomenological differences in the impact that the two scalar fields considered here, as well as the one studied in~\cite{bachsesteban2024}, have on WDs while also confirming the expected similarities between different mechanisms. These similarities arise from shared theoretical elements: the symmetron and dilaton mechanisms share an environment-dependent coupling, while dilaton and chameleon fields both feature a runaway potential. 

Future extensions of this work could incorporate more realistic modelling of the EoS for WDs, including leading-order Coulomb interactions, finite-temperature corrections, and possible contributions from the stellar envelope. A fully consistent treatment would also require WD evolution simulations from their progenitors within modified gravity frameworks---an essential ingredient that is currently unavailable (see~\cite{Saltas_2018} and references therein). A meaningful comparison with observational data becomes feasible only once key systematics are properly controlled. In particular, independent measurements of WD mass, radius, and effective temperature---such as those provided by eclipsing binaries---would be crucial to reduce degeneracies and observational uncertainties. These advances could pave the way for robust tests of the screening scenarios explored in this work. Additionally, it would be interesting to study radial and non-radial pulsation modes of symmetron- and dilaton-screened WDs. Asteroseismology techniques~\cite{Sotani_2005} could offer a powerful probe of the corresponding parameter spaces, which remain significantly less constrained than, for instance, those of the chameleon mechanism~\cite{Fischer:2024eic}.

\vspace{6pt} 





\authorcontributions{Conceptualisation, J.B.-E., I.L. and J.R.; methodology, J.B.-E.; software, J.B.-E.; validation, J.B.-E., I.L. and J.R.; formal analysis, J.B.-E.; investigation, J.B.-E.; resources, I.L. and J.R.; data curation, J.B.-E.; writing---original draft preparation, J.B.-E.; writing---review and editing, I.L. and J.R.; visualisation, J.B.-E.; supervision, I.L. and J.R.; project administration, I.L. and J.R.; funding acquisition, J.B.-E., I.L. and J.R. All authors have read and agreed to the published version of \mbox{the manuscript.}}

\funding{{J.B.-E.} 
 and I.L. acknowledge the Fundação para a Ciência e a Tecnologia (FCT), Portugal, for the financial support to the Center for Astrophysics and Gravitation---CENTRA, Instituto Superior Técnico, Universidade de Lisboa, through Project No. UIDB/00099/2020.
J.B.-E. is grateful for the support of this agency through grant No. SFRH/BD/150989/2021 in the framework of the IDPASC-Portugal Doctoral Program. 
I.L. also acknowledges FCT for the financial support through grant No. PTDC/FIS-AST/28920/2017.
J.R. is supported by a Ramón y Cajal contract of the Spanish Ministry of Science and Innovation with Ref.~RYC2020-028870-I. This work was supported by the project PID2022-139841NB-I00 of MICIU/AEI/10.13039/501100011033 and FEDER, UE.
This publication is based upon work from COST Action CA21136 Addressing observational tensions in cosmology with systematics and fundamental physics (CosmoVerse) supported by COST (European Cooperation in Science and Technology).}

\dataavailability{{Data are contained within the article.}
} 

\acknowledgments{The numerical part of this work has been performed with the support of the Infraestrutura Nacional de Computação Distribuída (INCD), funded by the FCT and FEDER under the project 01/SAICT/2016 nº 022153. J.B.-E. is grateful to the Universidad Complutense de Madrid for its hospitality during his visit between January and April 2025.}

\conflictsofinterest{The authors declare no conflicts of interest.}

\appendixtitles{yes} 
\appendixstart
\appendix
\section{Shooting Method Pseudocode}\label{sec:Pseudocode}

In this appendix, we have included a pseudocode that details the structure of our shooting method. As we explained in the main body of the paper, the purpose of this algorithm is to find the initial scalar field value, $\phi_0$, that leads to a final value $\phi(r_{\text{max}})$ matching the asymptotic boundary condition $\Bar{\phi}_\infty$ within a relative difference of $\phi_{\text{tol}}$. For both the symmetron and the dilaton, we only needed about 100 iterations to achieve convergence with a tolerance $\phi_{\text{tol}} = 10^{-10}$.

\begin{algorithm}[H]
\caption{Shooting method for Equations~\eqref{eq:background_Phi_Newton}--\eqref{eq:background_sigma_Newton} and the ODE system.} 
\begin{algorithmic}[1]
\State Initialize {$\phi_\text{try}$} with $\Tilde{\rho}_0$
\State Set {$\phi_{\text{var}}$} $\gets$ initial variation
\State Set counters: iteration $i \gets 0$, restart count $\gets 0$
\While{$i <$ max iterations}
    \If{{$\phi_\text{try}$} is invalid (negative or too large)}
        \State Adjust {$\phi_\text{try}$}, reduce {$\phi_{\text{var}}$}, increment restart count
        \State \textbf{continue}
    \EndIf

    \State Integrate ODE system with current {$\phi_\text{try}$}
    \If{solution is unphysical or diverges}
        \State Restart shooting with adjusted parameters
        \State \textbf{continue}
    \EndIf

    \State Compute $\phi_{\text{end}}=\phi(r_{\text{max}})$ and relative difference $(\Bar{\phi}_\infty - \phi_{\text{end}}) / \Bar{\phi}_\infty$
    \If{solution is best so far}
        \State Save current attempt
    \EndIf

    \If{converged to within tolerance $\phi_{\text{tol}}$ and scalar field is stable}
        \State \Return solution profile
    \EndIf

    \If{stagnation or slow convergence detected}
        \State Adjust {$\phi_\text{try}$}, $\phi_{\text{var}}$, and optionally $r_{\text{max}}$
        \State \textbf{continue}
    \EndIf

    \State Update {$\phi_\text{try}$} based on sign of the relative difference
    \State Increment iteration count $i \gets i + 1$
\EndWhile
\State \Return failure if maximum restarts reached
\end{algorithmic}
\end{algorithm}

\begin{adjustwidth}{-\extralength}{0cm}
\setenotez{list-name=Note}
\printendnotes[custom]

\reftitle{References}

\PublishersNote{}
\end{adjustwidth}
\end{document}